\newcommand{\ket}[1]{\ensuremath{\left|  #1 \right\rangle}}

\newcommand{\xx}{\mathbf{x}}
\newcommand{\yy}{\mathbf{y}}

\newcommand{\kk}{\mathbf{k}}
\newcommand{\pp}{\mathbf{p}}
\newcommand{\qq}{\mathbf{q}}

\newcommand{\epsk}{\epsilon_{\mathbf{k}}}
\newcommand{\um}{\mu\mathrm{m}}
\newcommand{\Bog}{\mathrm{Bog}}
\newcommand{\nconeb}{\overline{nc_1}}
\newcommand{\Tb}{\overline{T}}
\newcommand{\qb}{\overline{q}}
\newcommand{\dd}{\text{d}}

\documentclass[aps,pra,reprint,superscriptaddress]{revtex4-1}

\usepackage{amsmath,amssymb,graphicx,color,verbatim,soul,braket}

\usepackage{hyperref}
\usepackage{soul}

\usepackage[colorinlistoftodos, shadow,textsize= footnotesize]{todonotes}

\begin{document}

\title{Condensation and thermalization of an easy-plane ferromagnet in a spinor Bose gas}

\author{Maximilian Pr\"ufer}
\email{spincondensate@matterwave.de}
\affiliation{Kirchhoff-Institut f\"ur Physik, Universit\"at Heidelberg, Im Neuenheimer Feld 227, 69120 Heidelberg, Germany}
\affiliation{Vienna Center for Quantum Science and Technology, Atominstitut, TU Wien, Stadionallee 2, 1020 Wien, Austria}
\author{Daniel Spitz}
\affiliation{Institut f\"ur Theoretische Physik, Universit\"at Heidelberg, Philosophenweg 16, 69120 Heidelberg, Germany}
\author{Stefan Lannig}\affiliation{Kirchhoff-Institut f\"ur Physik, Universit\"at Heidelberg, Im Neuenheimer Feld 227, 69120 Heidelberg, Germany}
\author{Helmut Strobel}\affiliation{Kirchhoff-Institut f\"ur Physik, Universit\"at Heidelberg, Im Neuenheimer Feld 227, 69120 Heidelberg, Germany}
\author{J\"urgen Berges}
\affiliation{Institut f\"ur Theoretische Physik, Universit\"at Heidelberg, Philosophenweg 16, 69120 Heidelberg, Germany}
\author{Markus K. Oberthaler}\affiliation{Kirchhoff-Institut f\"ur Physik, Universit\"at Heidelberg, Im Neuenheimer Feld 227, 69120 Heidelberg, Germany}

\date{\today}

\begin{abstract}
	The extensive control of spin makes spintronics a promising candidate for future scalable quantum devices \cite{zutic_spintronics_2004}.
	For the generation of spin-superfluid systems \cite{sonin_spin_2010}, a detailed understanding of the build-up of coherence and relaxation is necessary.
	However, to determine the relevant parameters for robust coherence properties and faithfully witnessing  thermalization, the direct access to space- and time-resolved spin observables is needed.
	Here, we study the thermalization of an easy-plane ferromagnet employing a homogeneous one-dimensional spinor Bose gas \cite{stamper-kurn_spinor_2013,sadler_spontaneous_2006}.
	Building on the pristine control of preparation and readout \cite{kunkel_simultaneous_2019} we demonstrate the dynamic emergence of long-range coherence \cite{glauber_quantum_1963} for the spin field and verify spin-superfluidity by experimentally testing Landau's criterion \cite{landau_theory_1941}.
	We reveal the structure of the emergent quasi-particles: one `massive´ (Higgs) mode, and two `massless´ (Goldstone) modes -  a consequence of explicit and spontaneous symmetry breaking, respectively. 
	Our experiments allow for the first time to observe the thermalization of an easy-plane ferromagnetic Bose gas; we find agreement for the relevant momentum-resolved observables with a thermal prediction obtained from an underlying microscopic model within the Bogoliubov approximation \cite{uchino_bogoliubov_2010,phuc_effects_2011,symes_static_2014}. 
	Our methods and results pave the way towards a quantitative understanding of condensation  dynamics in large magnetic spin systems and  the study of the role of entanglement and topological excitations for its thermalization.
\end{abstract}
	\maketitle

\begin{figure}[]
	\linespread{1}
	\centering
	\includegraphics[width = \columnwidth]{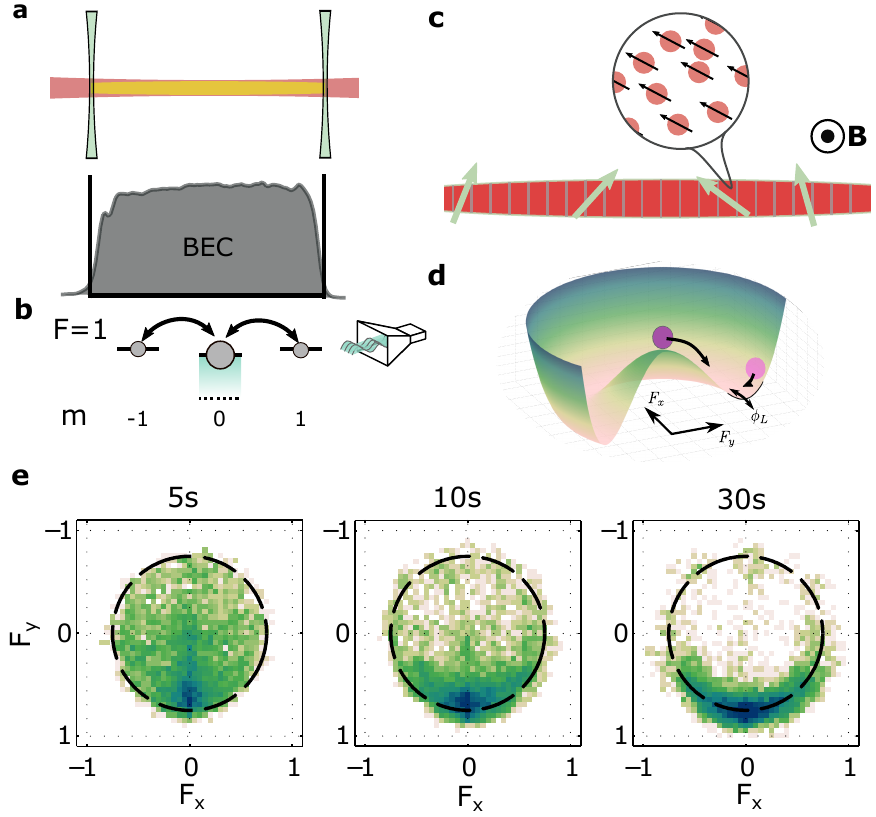} 
	\caption{\textbf{Homogeneous spinor Bose gas and easy-plane ferromagnetic properties. a,} We realize a homogeneous spinor BEC of $^{87}$Rb in a box-like trapping potential  by a combination of an elongated attractive potential (red) and two repulsive end caps (green; see Methods for details). The total density (grey shading) is flat over the extent of the cloud. \textbf{b,} Level structure of the $F = 1$ hyperfine manifold. We control the offset energy between the $m$-states by microwave dressing (blue shading) such that the system features easy-plane ferromagnetic properties in its ground state. \textbf{c,}  The spatial degree of freedom is continuous, however, in the analysis discretized by the finite pixel size of the camera and the imaging resolution ($\approx 1.2\,\mu$m). Each imaging volume (boxes) contains $\approx 500$ atoms which are described by continuous fields for density and spin. The spins orient themselves in the (easy-)plane orthogonal to the external magnetic field $\textbf{B}$. \textbf{d,} The transversal spin features two different types of excitations: A Goldstone mode and a  Higgs mode related to the excitation of the orientation $\phi_\text{L}$ and length $|F_\perp|$, respectively.  \textbf{e,} Histogram of the local spin normalized by the atom number, combining all spatial points and experimental realizations. In every realization the phase of the central spatial point is subtracted. The dashed line indicates $|F_\perp|=0.75$.
	}
	\label{Distribution}
\end{figure}

In recent years	analog quantum simulators with ultracold atoms allowed for unprecedented insights by implementing building blocks of complex condensed matter systems \cite{bloch_quantum_2012,gross_quantum_2017}. 
This opens up new possibilities for studying pressing questions concerning quantum many-body dynamics and thermalization \cite{rigol_thermalization_2008,eisert_quantum_2015,reimann_typical_2016,gogolin_equilibration_2016,mori_thermalization_2018,kaufman_quantum_2016,neill_ergodic_2016,clos_time-resolved_2016,evrard_many-body_2021}.
 For probing these phenomena in macroscopic systems often either the timescales are too short or the control to extract information is not given such that direct observation of the dynamical processes is not possible. 
 
Many dynamical phenomena emerging in the many-body limit, such as the build-up of long-range coherence, superfluidity or spontaneous symmetry breaking, can be studied in Bose-Einstein condensates (BEC) \cite{griffin_bose-einstein_1995}.
Here, the macroscopic occupation of the ground state together with a spontaneously broken symmetry manifests itself in a globally well-defined phase of the complex-valued order parameter in each realization.
This phase can be probed experimentally by interferometric measurements, which has been demonstrated with different platforms \cite{bloch_measurement_2000,deng_spatial_2007,damm_first-order_2017}.
In an easy-plane ferromagnetic system the order parameter is characterized by a well-defined magnitude in the transversal plane whereas all orientations in the plane are equally likely.
Theoretically, this is due to a spatial anisotropy, breaking the full rotational SO(3) symmetric part of the Hamiltonian down to a transversal SO(2) symmetry. 
In condensed matter physics prototype models include the XXZ model \cite{proukakis_bose-einstein_2017-2}, which recently has also been realized with ultracold atoms in lattice systems \cite{jepsen_spin_2020} and  Rydberg atoms \cite{geier2021floquet,scholl_microwave-engineering_2022}.

Here, we realize a spinor BEC of $^{87}$Rb with easy-plane ferromagnetic properties \cite{stamper-kurn_spinor_2013,uchino_spinor_2015} in a quasi-one-dimensional box trap \cite{navon_quantum_2021} (see Fig. 1a).
It consists of three internal states, labelled by their magnetic quantum number $m \in \{0,\pm1\}$. The system features rotationally invariant ferromagnetic spin-spin interactions described by $\hat{	\mathcal{H}}_{{s}} = c_1 \int \dd V\, \hat{\boldsymbol{F}}^2/2$, where $c_1<0$ is the spin-spin interaction constant and $\hat{\boldsymbol{F}}$ denotes the spin operator (see Methods for details).
A quadratic Zeeman shift $q$ induced by the magnetic field plays the role of the isotropy-breaking term; it shifts the energy of the $m=\pm1$ levels (see Fig. 1b) and is explicitly given by ${\hat{\mathcal{H}}_q=q \int \dd V\, (\hat{N}_{+1}+\hat{N}_{-1})}$. 
We adjust $q$ by using off-resonant microwave dressing \cite{gerbier_resonant_2006} such that the mean-field ground-state exhibits easy-plane ferromagnetic properties ($0< q  < 2n |c_1|$; $n$ is the atomic density) and our initial conditions restrict the dynamics to the spatially averaged longitudinal ($z$-) spin being zero.
In addition to the spin interactions, our system exhibits SO(3)-invariant density-density interactions described by $\hat{	\mathcal{H}} _{{d}}= c_0\int \dd V\, \hat{N}^2/2$, with interaction constant $c_0$ and $|c_0/c_1|\approx200$ \cite{stamper-kurn_spinor_2013}.

The capability to extract the relevant order-parameter field \cite{kunkel_simultaneous_2019} allows us to study  the build-up of long-range coherence in a time- and space-resolved fashion; accessing the full structure factor of the observables defining the Hamiltonian is the handle to faithfully witness thermalization.
We experimentally examine the order parameter, which is the transversal spin degree of freedom, by acquiring many realizations of the complex-valued field $F_\perp (y)= F_x (y) +\text{i}F_y (y)$ using spatially resolved joint measurements based on positive operator valued measures (POVM) \cite{kunkel_simultaneous_2019,prufer_experimental_2020}.
We obtain a value for the  transversal spin $F_\perp(y)=|F_\perp| e^{-i\phi_L}$ with length $|F_\perp|$ and orientation in the plane $\phi_L$.
The position $y$ {along the long axis of the cloud } is discretized by our imaging resolution; in each typical imaging volume we infer the spin from an average over $\sim 500$ atoms which are described by a spin field, i.e.\ taking nearly continuous values (see Fig.\,1b), which we identify as the macroscopic order-parameter field describing the spin condensation.

For studying the condensation dynamics, we initialize the system far from equilibrium without well-defined spin length, and fluctuations solely in the plane.
We visualize the emergence of a spin ($F_\perp$) field by evaluating the histogram of $F_\perp$ taking into account all spatial positions and realizations (see Fig.\,1e).
After 5\,s, which corresponds to $\approx10\times$ the typical time scale of the spin interaction energy $t_s = h/(n|c_1|)$, the spin is still far from equilibrium and shows large fluctuations in orientation and length.
After 30\,s ($\approx 60\times t_s$) of evolution time we find that the fluctuations settle around a well-defined spin length $|F_\perp|$ and the phase $\phi_\text{L}$ becomes well-defined over the whole sample, i.e.\ long-range order emerges. This is expected for a thermal state incorporating spontaneous symmetry breaking in the transversal spin degree of freedom and can be intuitively grasped by looking at the underlying mexican-hat-like free-energy potential (see Fig.\,1d and \cite{kawaguchi_spinor_2012}).

\begin{figure}
	\linespread{1}
	\centering
	\includegraphics[width = \columnwidth]{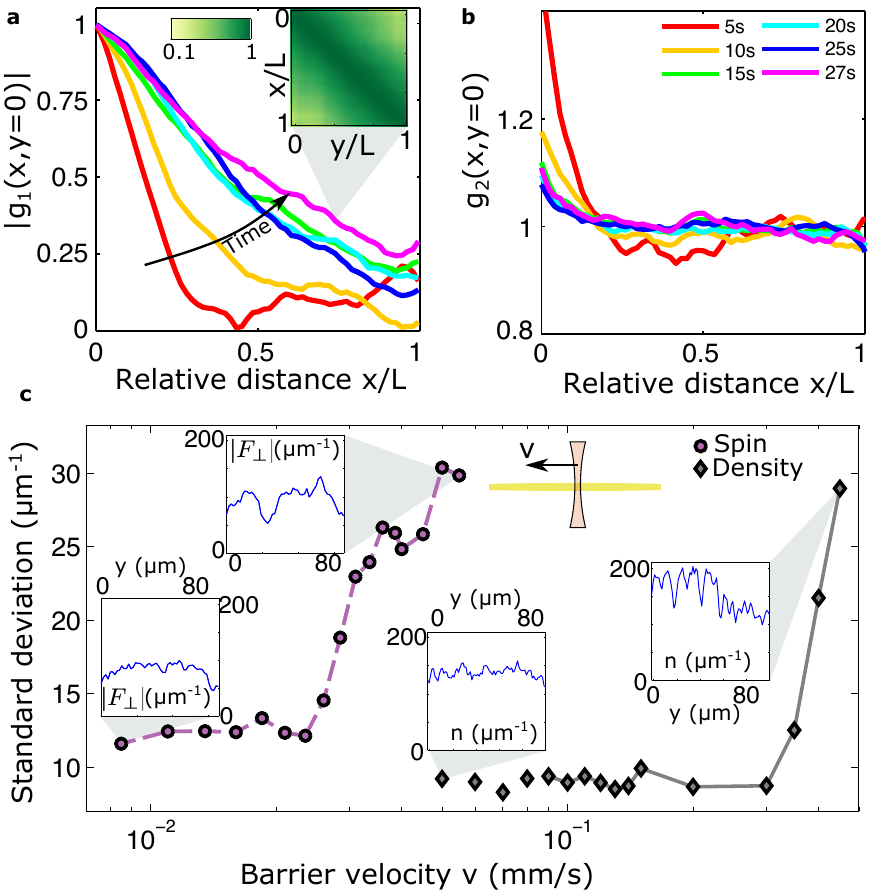} 
	\caption{\textbf{Emergence of long-range coherence and superfluidity.}  \textbf{a,} Absolute value of first order coherence $|g_1 (x, y=0)|$ of transversal spin $F_\perp$; reference point $(y=0)$ is chosen at the left edge of the cloud with system size $L=74\,\mu\text{m}$. We observe a build-up of long-range order, i.e.\ for long times the system features non-zero coherence over its whole size. (Inset) Two-dimensional coherence function $|g_1 (x,y)|$ after 27\,s evolution time. For long times we find the correlations to be translation-invariant.
		\textbf{b,} Second order coherence of the transversal spin showing the evolution and character of spin length fluctuations. 
		\textbf{c,} Superfluid properties of the spin condensate. Standard deviation along the cloud of spin length (purple) and density (grey) for different speeds $v$ of the local perturbation. The rapid increase at finite speed indicates superfluid properties of spin and density. Insets show representative single {realizations} of the spin length and total density in the different regimes.
	  } 
	\label{Distribution}
\end{figure}

\begin{figure*}
	\linespread{1}
	\centering
	\includegraphics[width = 2\columnwidth]{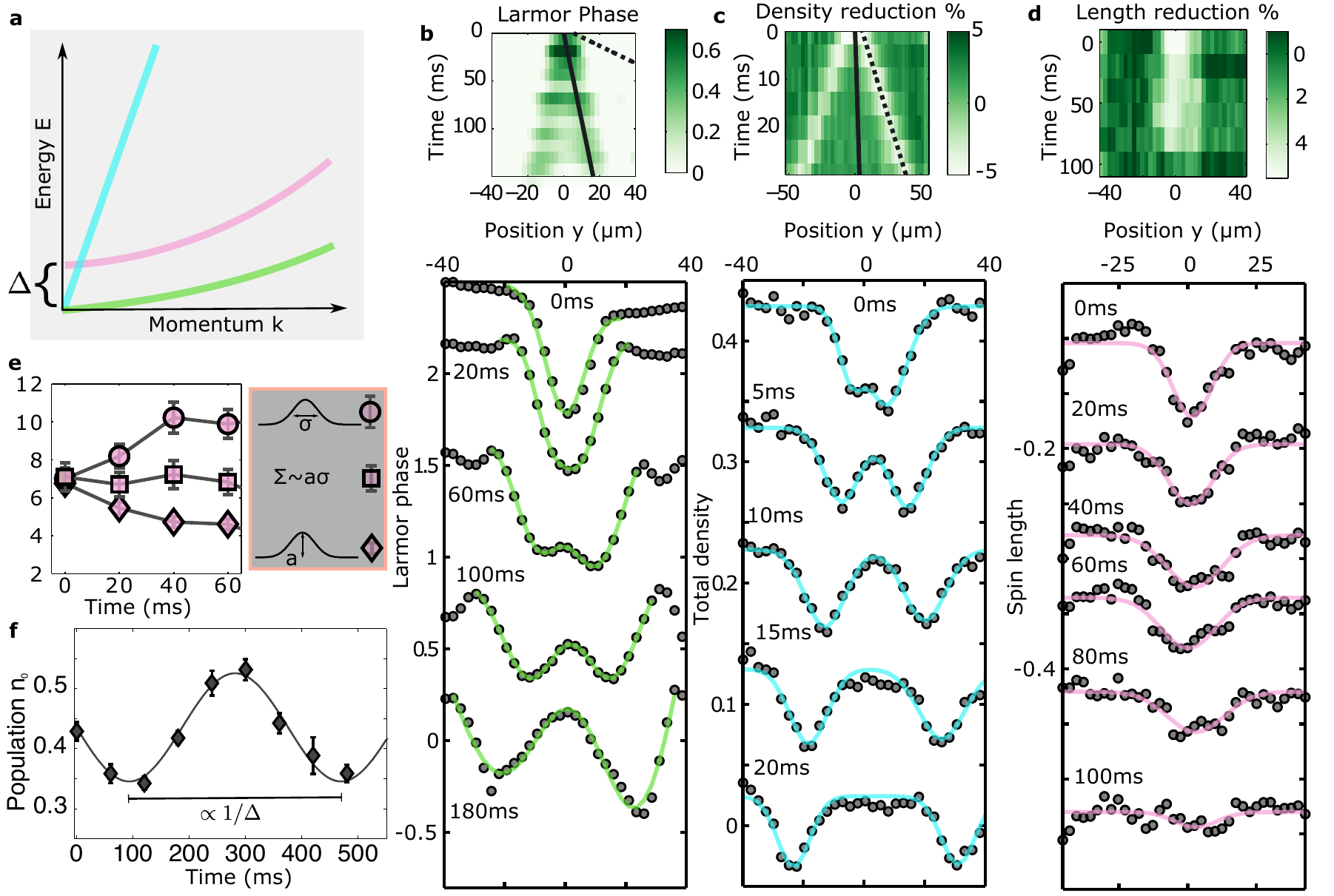} 
	\caption{\textbf{Local spin and density control enables probing of quasi-particle properties.}  \textbf{a,} Schematics of the three Bogoliubov dispersion relations. 	
		\textbf{b-d,} Time evolution of orientation (\textbf{b,}), total density (\textbf{c,}) and spin length (\textbf{d,}) after local perturbation of the spin condensate. The upper panels show all evolution times and lower panels selected 1D cuts.
		We find a splitting of the imprinted wavepacket for phase and total density according to the expected linear dispersion relations (green and blue); solid lines are Gaussian fits. Strikingly, the speed of sound differs by one order of magnitude reflecting the energy scales (solid black line corresponds to $v = 110\mu\text{m} / \text{s}$ and dashed line to $v = 1100\mu\text{m} / \text{s}$). In contrast, the spin length excitation disperses.  \textbf{e,} Results of Gaussian fits to spin length excitation. $1\sigma$ width (circles), the amplitude $a$ (diamonds) and the integral $\propto a\sigma$ (squares) are shown. We find an increasing (decreasing) width (amplitude) while the integral stays nearly constant; this is in accordance with an underlying gapped quadratic dispersion.
		\textbf{f,} Oscillation of the $m = 0$ population after perturbing the $k=0$  mode of the spin length. The oscillation frequency is a measure of the gap $\Delta$ of the quadratic mode identified in \textbf{d}. All shown error bars are 1 s.d.\ of the mean.}
	\label{Distribution}
\end{figure*}

\begin{figure*}
	\linespread{1}
	\centering
	\includegraphics[width = 2\columnwidth]{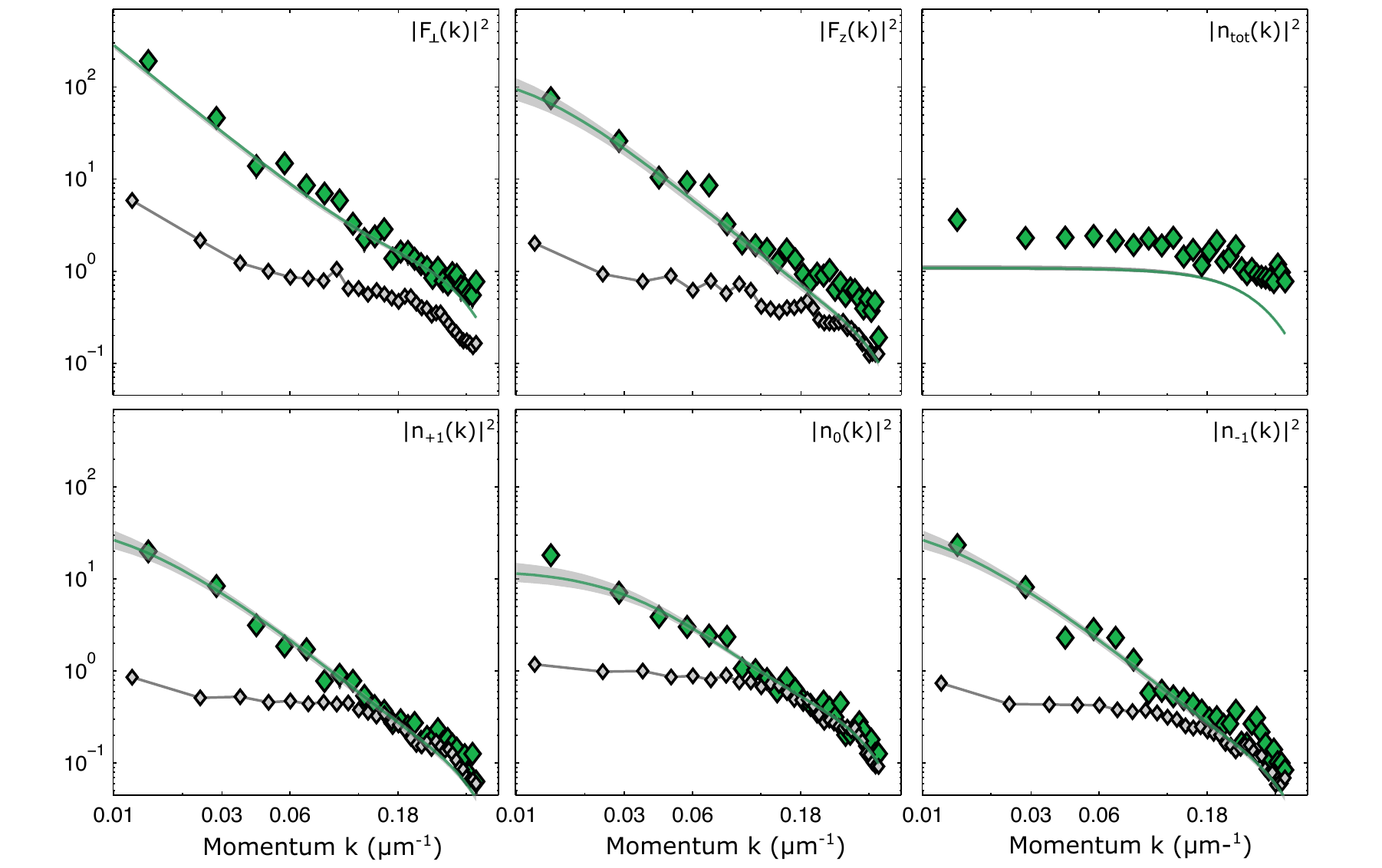} 
	\caption{\textbf{Structure factors of different observables at late times.} We show experimental structure factors (green diamonds; error bars smaller than plot marker) after 30\,s evolution time where the system behaves stationary. Experimental uncertainties are smaller than marker size. We compare to a thermal prediction within number-conserving Bogoliubov theory  (green line; grey band indicates 68\% confidence interval of statistical and systematic uncertainties; for details on the fit and parameters see Methods). Grey diamonds indicate reference noise level of a coherent spin state prepared from a single component gas by a global spin rotation.}
	\label{Distribution}
\end{figure*}

To test for eventual spin condensation, we characterize the coherence properties of the transversal spin by evaluating first and second order coherence functions \cite{glauber_quantum_1963} with $g_1(x,y) \propto {\braket{{\hat{F}}_\perp^\dagger(x){\hat{F}}_\perp(y)}}$ and ${g_2(x,y) \propto {\braket{{\hat{F}}_\perp^\dagger(x){\hat{F}}^\dagger_\perp(y){\hat{F}}_\perp(x){\hat{F}}_\perp(y)}}}$, respectively (see Methods for details). 
In contrast to earlier experiments observing the emergence of long-range coherence in one-component BECs \cite{bloch_measurement_2000,kohl_growth_2002,ritter_observing_2007}, we do not rely on spatial interference as we access the relevant spin field directly by joint measurements entailing interference in the internal degrees of freedom \cite{kunkel_simultaneous_2019,kunkel_detecting_2022}.
We find that coherence is built up dynamically and the system finally features long-range order, i.e.\ non-zero $|g_1(x,y)|$ over the whole extent of the atomic cloud (see Fig.\,2a).
At the same time the spin length fluctuations, quantified by $g_2(x,y)$ (see Fig.\,2b), settle close to unity at zero distance as expected for a weakly interacting Bose-Einstein condensate \cite{naraschewski_spatial_1999,ottl_correlations_2005,hodgman_direct_2011,perrin_hanbury_2012,cayla_hanbury-brown_2020}.

To characterize the final state, we first test for superfluidity of the spin as well as the density; in the spirit of Landau \cite{landau_theory_1941}, we drag a well-localised obstacle (see Methods) coupling to density and spin through the BEC \cite{raman_evidence_1999,weimer_critical_2015,kim_observation_2020} and measure the response of the system.
We quantify the response by evaluating the mean standard deviation of the total density and the transversal spin length along the cloud.
The breakdown of superfluidity is signalled by a rapid increase of the response at a non-zero critical velocity.
We find two different critical velocities for spin and density (see Fig.\,2c).
While the spin shows superfluidity up to $v_{c,s} \simeq 3\times 10^{-2}\,$mm/s, the density tolerates a moving barrier for up to 10 times faster speeds. 
This is consistent with the interaction strengths and the corresponding stiffness of the degrees of freedom.

In the following we address the underlying structure in more detail.
With two spontaneously broken symmetries,  the U(1) symmetry of the total density and the SO(2) symmetry of the spin orientation, we anticipate two Goldstone-like modes with linear dispersions in the infrared.
The different energy scales of density and spin interactions are reflected in two associated sound speeds; they are theoretically expected to differ by more than an order of magnitude which is consistent with the observed critical velocities.
Additionally, the symmetry explicitly broken by $\hat{\mathcal{H}}_s+\hat{\mathcal{H}}_q$  leads to a Higgs-like gapped mode (see Fig.\,3a).
Compared to two-component BECs we find an additional mode due to the increased number of degrees of freedom \cite{recati_breaking_2019,cominotti_observation_2021}.

Experimentally, we probe the three different modes by applying local perturbations (see Fig.\,3 and Methods for details).
After the perturbation we observe and analyze the temporal evolution to learn about the underlying structure of the dispersion relations.
First, we probe the linear mode associated with the spin orientation by  imprinting a spatially varying orientation $\phi_L$ pattern onto the thermalized state.
The probing scheme is based on our capabilities to combine global and local radio frequency spin rotations with fixed relative phases.
The initially imprinted Gaussian wavepacket  splits up into two wavepackets travelling with velocities of $\pm v_{{s}}$ -- a clear indication for a linear dispersion relation.
To access the density degree of freedom we imprint a Gaussian-shaped density reduction of $\sim 5\%$ and observe again a splitting of the initially prepared wave packet.
The difference in energy scales of density and spin is reflected in the two observed sound speeds $v_{{d}}$ and $v_{{s}}$ which we find to be $v_{{d}}/v_{{s}}  = 10 \approx \sqrt{|c_0/c_1|}$.
Finally, the gapped mode is associated with excitations of the spin length which we perturb with a Gaussian length modulation.
Strikingly, we find no splitting but a decaying amplitude and growing width of the prepared perturbation.
Additionally we measure the gap of this mode; it manifests as a finite oscillation frequency when exciting the $k = 0$ mode (see Fig.\,3f).
We find a $q$-dependent, non-zero oscillation frequency consistent with expectations concerning the nature of the underlying Bogoliubov mode in the easy-plane ferromagnetic phase (see Ext.Data.Fig.1 and Methods for details).

We now turn to the question if we experimentally realized a thermal ensemble; to ease its realization we prepare an elongated spin initial condition and evolve it for 30\,s.
Experimentally, we extensively characterize the thermalized state of our system by using different structure factors, such as those of the spin in transversal as well as in longitudinal ($z$-) direction and the densities of the three $m$-components.
{To separate the single components we employ a Stern-Gerlach magnetic field gradient and a short time-of-flight (2\,ms).
We are able to extract momentum-resolved structure factors by Fourier-transforming the spatial profiles.
Special care has to be taken to measure the total density fluctuations since phase fluctuations are transformed into density fluctuations during any time-of-flight \cite{betz_two-point_2011}.
Therefore we employ an in-situ imaging of the total density to access the low level of fluctuations which is in our regime of the order of atom shot-noise (see Methods for details).}

The structure factor (see Fig.\,4) as well as the local fluctuations (see Ext.\,Data Fig.\,2) of all observables are consistently described using a thermal prediction.
The latter is obtained for the spinor Bose gas with contact interactions within the Bogoliubov approximation \cite{bogolubov_theory_nodate,uchino_bogoliubov_2010} using a single temperature for all three quasiparticle modes.
We thus conclude that the system has evolved  to a thermal state within experimentally accessible timescales.
The found temperature is  $(57 \pm 3)\,$Hz ($\sim 3\,$nK) and thus a factor of $\approx5 \times$ smaller than the density-density interaction ($nc_0 = 252 \pm54\,$Hz) and more than one order of magnitude larger than the spin-spin interaction energy scale ($|nc_1|= 1.17 \pm0.25\,$Hz).
The temperature is consistent with theoretical estimates for the critical temperature for the emergence of the easy-plane ferromagnetic phase \cite{phuc_effects_2011,kawaguchi_finite-temperature_2012}.
Interestingly, the single $m$-densities feature high fluctuations also in comparison with three independent single-component condensates at corresponding density, interaction and temperature.

We repeat our measurement close to the phase boundary at $q=0$ and find structures beyond thermal Bogoliubov theory in that case (see Ext.Data Fig.\,3).
The observed enhanced fluctuations can be associated with long-lived non-linear excitations of the spin \cite{blakie_solitons_2022} which are energetically less suppressed for lower $q$. 

In conclusion, the high degree of control allows us to experimentally observe the thermalization process of an easy-plane ferromagnet.
This sets the foundations for studies in quantum field settings addressing the microscopic processes for thermalization as well as its absence due to e.g.\ long-lived topological defects.
The robust generation of a spin superfluid is a prerequisite for spin Josephson junctions where finite temperature effects and spin-density separation can now be studied on a new quantitative level due to the direct access to the order parameter.
\,

\noindent\textbf{Acknowledgements}\\
We thank Jan Dreher for experimental assistance.\\
This work is supported by ERC Advanced Grant Horizon 2020 EntangleGen (Project-ID 694561), the Deutsche Forschungsgemeinschaft (DFG, German Research Foundation) under Germany’s Excellence Strategy EXC2181/1-390900948 (the Heidelberg STRUCTURES Excellence Cluster) and the Collaborative Research Center - Project-ID 27381115 - SFB 1225 ISOQUANT.  M.P. has received funding from the European Union’s Horizon 2020 research and innovation programme under the Marie Skłodowska-Curie grant agreement No 101032523.\\
\noindent\textbf{Author contributions}, 
M.P., S.L. and H.S. took the measurement data. M.P., D.S., S.L., H.S. and M.K.O. discussed the measurement results and analysed the data. D.S., M.P.,  and J.B. elaborated the theoretical framework. All authors contributed to the discussion of the results and the writing of the manuscript.

\noindent\textbf{Competing financial interests}\\
The authors declare no competing financial interests.

\noindent\textbf{Data availability statement}\\
Source data and all other data that support the plots within this paper and other findings of this study are available from the corresponding author upon reasonable request.

\,
\onecolumngrid
\newpage

\begin{center}
	\textbf{Methods}
\end{center}

\subsection{Experimental details}

We prepare a spinor Bose-Einstein condensate of $^{87}$Rb in a quasi-one-dimensional trapping geometry. 
Details concerning the preparation and readout of the transversal spin can be found in \cite{prufer_observation_2018,prufer_experimental_2020,kunkel_simultaneous_2019,kunkel_detecting_2022}

Here, we employ a box-like trapping potential. 
We use a weakly focused {red-detuned} laser beam {creating to a quasi-one-dimensional trapping potential with $\omega_l \approx 2\pi\times 1.7\,$Hz and $\omega_r \approx 2\pi\times 170\,$Hz; repulsive potential walls are created by}  two blue-detuned laser beams {which results in a trapping volume of adjustable size around the centre of the harmonic trap.} 
The longitudinal harmonic potential is in {good approximation} constant over the employed sizes and, thus, effectively leads to a 1D box-like confinement for the atomic cloud.
For the measurements of the thermalized state shown in Fig.\,4 we utilize a box size of $\sim100\,\mu$m.

\textit{Initial conditions:}
For the detailed observation of the emergence of coherence we prepare the atoms in the state $\ket{F,m} = \ket{1,0}$, the so-called polar state.
To allow for thermalization {for shorter times}, we initially prepare a coherent spin state with maximal length.
For this we apply a $\pi/2$-rf rotation with the atoms initially prepared in the state $\ket{F,m} = \ket{1,-1}$.
{As a reference noise level (grey diamonds in Fig.\,4 and Ext.\,Data Fig.\,3) for the thermalized structure factor, we prepare this coherent spin state by performing the rotation after holding the atoms in $\ket{1,-1}$ for 30\,s.}

\textit{Readout: }After the evolution time $t$ we image the atomic densities using spatially resolved absorption imaging.
Employing a Stern-Gerlach magnetic field gradient followed by a short time-of-flight (TOF; $2\,$ms), we are able to image {the atomic densities of} all 8 magnetic sublevels of the $F = 1$ and $ F=2$ hyperfine manifolds of the electronic ground state.
{Additional coherent microwave and radio-frequency manipulations before the imaging allow us to  map the two spin projections, $F_x$ and $F_y$, of the transversal spin \cite{prufer_experimental_2020} onto measureable densities}. 
With this we infer the complex-valued transversal spin $F_\perp(y) = F_x(y)+iF_y(y) =  |F_\perp|(y)\text{e}^{-i\phi_\text{L}(y)}$ as a function of position $y$.  
The position $y$ is the centre of a spatial bin which contains $\sim500$ atoms {and has a spatial extension of $\sim1.2\,\mu$m along the cloud (we bin three adjacent camera pixels where each pixel corresponds to $420\,$nm in the atom plane)}.

{For the measurement of the density fluctuations $|N_{\text{tot}}|^2(k)$ we take in-situ images without spin resolution (without Stern-Gerlach separation). This is important since any free propagation will transform phase fluctuation to density fluctuations leading to strongly enhanced structure factor \cite{manz_two-point_2010}. It is important to note that the observed increased fluctuations compared to the spin coherent state by a factor of two can be a result of only one particle per k-mode. For the spin observables and the single densities we checked that the enhanced fluctuations due to the TOF are negligible.}

\subsection{Local perturbation}
To access the superfluid properties of the spin and density degrees of freedom we use a localized perturbation (r.m.s width $ \sim5\,\mu$m) that we drag through the thermalized system.
Specifically we use a blue detuned, steerable laser beam (760\,nm) which position is controlled by an acusto-optical deflector.
Using a linear frequency ramp we implement a sweep over the cloud with fixed velocity which we change over two orders of magnitude.
The density is probed after 35\,s and the spin after 20\,s evolution time.
{The ramp duration for the lowest speed is $\approx 18\,$s.}

 \textit{Local perturbation of the Larmor phase:} We use a combination of global and local rf rotations (see \cite{lannig_collisions_2020} for details on local rf rotations). 
A first global $\pi/2$-rf rotation around the $x$-axis maps the $z$-axis onto the $y$-axis. 
Using a local rotation with a well-defined phase with respect to the global rotation we perform a rotation with variable angle around the $y$-axis. 
Because of the performed mapping this effectively leads to rotation around the $z$-axis in the original coordinate system. 
At time $\Delta\tau = 210\,\mu$s after the first global rf pulse we apply a global rf $\pi$-pulse followed by a second global rf $\pi$/2-pulse after another time delay of $\Delta\tau$, where all pulses rotate around the same axis. 
This constitutes a spin echo sequence which additionally executes a full $2\pi$ spin rotation which ensures that the global rotation pulses do not excite the system. 
The last $\pi$/2-pulse maps the local rotation axis back to the z-axis in the original system.
The perturbation has an approximate Gaussian shape with a r.\,m.\,s.\,width of $\sim 5\,\mu$m according to the shape of the used laser beam.

 \textit{Local perturbation of the total density:} We reduce the total density locally by $\approx 5\%$ by shining a blue detuned laser beam ($760\,$nm) onto the centre of the cloud. 
We adiabatically ramp up the potential in 100\,ms such that we get no further excitations in the density; after the ramp the potential is instantaneously switched off to generate the wavepacket.

 \textit{Local perturbation of the transversal spin length $|F_\perp|$:} We induce a local density reduction by applying the same blue-detuned laser beam. 
During the evolution time of 30\,s we let the system thermalize subject to the local density reduction.
This effectively leads to a spatially dependent mean-field ground state spin length.
We linearly ramp down the potential over 50\,ms; this implements an adiabatic ramp for the total density and a rapid switch off for the spin.

For experimentally accessing the gap we excite the $k= 0$ mode of the spin length by changing the phase of the $m= 0$ component (spinor phase) globally.
For this we use two microwave $\pi$-pulses between $\ket{1,0}$ and $\ket{2,0}$, where the second pulse is phase shifted by $\Delta \phi$.
We record the subsequent oscillations of the $m = 0$ population and fit a sinusoidal function to extract the frequency.
The theoretical prediction for the gap $\Delta$, deduced from the oscillation, and the $m=0$ ground state population $n_0$ in the easy-plane ferromagnetic phase is \cite{uchino_bogoliubov_2010}:
\begin{equation}
\Delta = \sqrt{4n^2c_1^2-q^2}    \,\,\,\,\,\,\,\,\,\,\,\,\,\,\,\,\,\,\,\,\, \text{and} \,\,\,\,\,\,\,\,\,\,\,\,\,\,\,\,\,\,\,\,\,	n_0 = \frac{1}{2} - \frac{q}{4nc_1}\,.
\end{equation}

Assuming $n_{+1} = n_{-1}$ these formulae hold also true for $0>q>2nc_1$ (dashed lines in Ext.\,Data Fig.\,1).

\subsection{Experimental coherence functions and structure factor}

In every experimental realization $(i)$ we measure atomic densities from which we infer single shot realizations $O^{(i)}$ of different observables $\hat{O}$.
The quantum expectation value is approximated by averaging over many realizations as 

\begin{equation}
O = \braket{\hat{O}} =  \frac{1}{N_S} \sum_{i  = 1}^{N_S} O^{(i)}\,,
\end{equation}
where $N_S$ is the number of realizations.

The coherence functions of the transversal spin are explicitly given by:
\begin{equation}
	g_1(x,y) = \frac{\braket{\hat{F}_\perp^\dagger(x)\hat{F}_\perp(y)}}{\sqrt{\braket{{\hat{F}}_\perp^\dagger(x){\hat{F}}_\perp(x)}\braket{{\hat{F}}_\perp^\dagger(y){\hat{F}}_\perp(y)}}}
\end{equation}
and
\begin{equation}
	g_2(x,y) = \frac{{\braket{{\hat{F}}_\perp^\dagger(x){\hat{F}}^\dagger_\perp(y){\hat{F}}_\perp(x){\hat{F}}_\perp(y)}}}{{\braket{{\hat{F}}_\perp^\dagger(x){\hat{F}}_\perp(x)}\braket{{\hat{F}}_\perp^\dagger(y){\hat{F}}_\perp(y)}}}\,.
\end{equation}
For the inferred single shot results of the transversal spin $F_\perp(x)$ the $^\dagger$ is treated as the complex conjugate.

The structure factors as a function of the spatial momentum $k$ are defined as

\begin{equation}
	\braket{|\hat{O}(k)|^2} = |O(k)|^2 =   \frac{1}{N_\text{tot}}  \frac{1}{N_S} \sum_{i  = 1}^{N_S} |DFT_{x\rightarrow k}  \left( O^{(i)}  (x) - {O  (x)}\right) |^2  \,,
\end{equation}
where $DFT_{x\rightarrow k}  $ is the discrete Fourier transform, $k=1/\lambda$ the spatial momentum.
All structure factors are normalized by the mean total atom number $N_\text{tot}$ to obtain an atom number independent measure for the fluctuations and allow comparison between theory and experiment;  for the total density structure factor a value of one corresponds to the atomic shot noise level.

\subsection{Bogoliubov transformations in the easy-plane ferromagnetic phase}\label{SecBogTrafosBAPhase}
We explicitely derive the Bogoliubov transformations in the easy-plane ferromagnetic phase {($0 <  q/(n|c_1|) < 2$)}. {Here we set $\hbar = 1$.}

In terms of total density and spin operators
\begin{equation}
\hat{N}(\mathbf{x}) = \sum_{m=-1}^1\hat{N}_m(\mathbf{x})  = \sum_{m=-1}^1 \hat{\psi}^\dagger_m(\mathbf{x})\hat{\psi}_m(\mathbf{x}),\qquad \hat{F}_\nu(\mathbf{x}) = \sum_{m,m'=-1}^1\hat{\psi}_m^\dagger(\mathbf{x})(f^\nu)_{mm'}\hat{\psi}_{m'}(\mathbf{x}),
\end{equation}
{with the} spin-1 matrices
\begin{equation}
f^x = \frac{1}{\sqrt{2}}\left(\begin{matrix}
0 & 1 & 0\\
1 & 0 & 1\\
0 & 1 & 0
\end{matrix}\right),\qquad f^y = \frac{i}{\sqrt{2}}\left(\begin{matrix}
0 & -1 & 0\\
1 & 0 & -1\\
0 & 1 & 0
\end{matrix}\right),\qquad f^z = \left(\begin{matrix}
1 & 0 & 0\\
0 & 0 & 0\\
0 & 0 & -1
\end{matrix}\right),
\end{equation}
the system Hamiltonian reads 
\begin{equation}
\hat{\mathcal{H}} = \int d^3\xx \bigg[\sum_{m=-1}^1 \hat{\psi}_m^\dagger(\mathbf{x})\bigg(-\frac{\nabla^2}{2M} + q m^2\bigg)\hat{\psi}_m(\mathbf{x}) + \frac{c_0}{2} :\hat{N}^2(\mathbf{x}): + \frac{c_1}{2}\sum_{\nu=x,y,z} :\hat{F}_\nu^2(\mathbf{x}): \bigg].
\end{equation}
With momentum-space creation and annihilation operators
\begin{equation}
\hat{a}_{\kk,m}^\dagger = \frac{1}{\sqrt{V}}\int d^3\xx\,  \hat{\psi}^\dagger_m(\xx)e^{i\kk\xx},\qquad  \hat{a}_{\kk,m} = \frac{1}{\sqrt{V}}\int d^3\xx\, \hat{\psi}_m(\xx) e^{-i\kk\xx},
\end{equation}
the Hamiltonian in the number-conserving Bogoliubov approximation becomes \cite{kawaguchi_spinor_2012}
\begin{align}
\hat{\mathcal{H}}_B = &\, E_0 + \sum_{\kk\neq \mathbf{0},m} (\epsilon_\kk + q m^2 - \mu) \hat{n}_{\kk,m}\nonumber\\
 & \, + \frac{N}{V}\sum_{j,j',m,m'} \sum_{\kk\neq \mathbf{0}} (\Gamma_{jj',m'm} + \Gamma_{jm,m'j'}) \zeta_{j'}\zeta_{m'}^* \hat{a}^\dagger_{\kk,j} \hat{a}_{\kk,m}\nonumber\\
 &\, + \frac{N}{2V} \sum_{j,j',m,m'}\sum_{\kk\neq\mathbf{0}} \Gamma_{jj',mm'}\big(\zeta^*_j\zeta^*_m\hat{a}_{-\kk,m'}\hat{a}_{\kk,j'} + \zeta_{m'}\zeta_{j'}\hat{a}^\dagger_{\kk,j}\hat{a}^\dagger_{-\kk,m}\big).\label{EqHamiltonianBogoliubov}
\end{align}
with $\epsk= \kk^2/(2M)${, atom mass $M$, total atom number $N$} and $\hat{n}_{\kk,m} = \hat{a}^\dagger_{\kk,m}\hat{a}_{\kk,m}$. {Theoretically, momentum $|\kk|$ corresponds to wavelength $\lambda = 2\pi/|\kk|$.} The spinor $(\zeta_m)$ specifies the normalized condensate configuration and we set $\hat{a}_{\kk=0,m}=\sqrt{N}\zeta_m$. $\Gamma_{jj',mm'}$ denotes density and spin interactions,
\begin{equation}
\Gamma_{jj',mm'}\equiv c_0 \delta_{jj'}\delta_{mm'} + c_1 \sum_{\nu=x,y,z} f^\nu_{jj'}f^\nu_{mm'}.
\end{equation}
The ground state energy is given by
\begin{equation}\label{EqGroundStateEnergy}
E_0\equiv N\bigg[ \sum_m q m^2 |\zeta_m|^2 + \frac{N-1}{2V} \sum_{j,j',m,m'} \Gamma_{jj',mm'} \zeta_j^*\zeta_m^*\zeta_{m'}\zeta_{j'}\bigg],
\end{equation}
the chemical potential reads
\begin{equation}\label{EqChemicalPotentialBogoliubov}
\mu \equiv \sum_m q m^2 |\zeta_m|^2 + \frac{2N-1}{2V} \sum_{j,j',m,m'}\Gamma_{jj',mm'}\zeta^*_j\zeta^*_m\zeta_{m'}\zeta_{j'}.
\end{equation}

We set $\sin \theta = \sqrt{1/2-q/(4n|c_1|)}$, such that the mean-field ground state reads $\zeta=(\sin\theta/\sqrt{2},\cos\theta,\sin\theta / \sqrt{2})$ \cite{uchino_bogoliubov_2010}. The initial orthogonal transformation, given by
\begin{equation}
\left(\begin{matrix}
\hat{a}_{\kk,d}\\
\hat{a}_{\kk,\theta}\\
\hat{a}_{\kk,f_z}
\end{matrix}\right) = \left(\begin{matrix}
\sin\theta / \sqrt{2} & \cos\theta & \sin\theta / \sqrt{2}\\
\cos\theta/\sqrt{2} & -\sin\theta & \cos\theta/\sqrt{2}\\
1/\sqrt{2} & 0 & -1/\sqrt{2}
\end{matrix}\right)\left(\begin{matrix}
\hat{a}_{\kk,1}\\
\hat{a}_{\kk,0}\\
\hat{a}_{\kk,-1}
\end{matrix}\right) \equiv A(\theta)\left(\begin{matrix}
\hat{a}_{\kk,1}\\
\hat{a}_{\kk,0}\\
\hat{a}_{\kk,-1}
\end{matrix}\right),
\end{equation}
leads to a description of the system in terms of longitudinal and transversal spin fluctuations.

The longitudinal ($z$-) spin fluctuations can be diagonalized using the Bogoliubov transformation \cite{uchino_bogoliubov_2010}
\begin{equation}
\hat{b}_{\kk,f_z} = u_{\kk,f_z} \hat{a}_{\kk,f_z} + v_{\kk,f_z} \hat{a}^\dagger_{-\kk,f_z},
\end{equation}
where
\begin{equation}
u_{\kk,f_z}\equiv \sqrt{\frac{\epsk+q/2+E_{\kk,f_z}}{2E_{\kk,f_z}}},\qquad v_{\kk,f_z}\equiv \sqrt{\frac{\epsk+q/2-E_{\kk,f_z}}{2E_{\kk,f_z}}},
\end{equation}
with the dispersion
\begin{equation}
E_{\kk,f_z} = \sqrt{\epsk (\epsk+q)}.
\end{equation}

To diagonalize transversal spin fluctuations we follow the diagonalization procedure outlined in \cite{kawaguchi_spinor_2012}. 
We obtain mode energies $\pm E_{\kk,+}$ and $\pm E_{\kk,-}$ as in \cite{kawaguchi_spinor_2012}, explicitely given by
\begin{equation}
E_{\kk,\pm} = \sqrt{\epsk^2+n(c_0-c_1)\epsk + 2n^2c_1(c_1-c_q)\pm E_1(\kk)},
\end{equation}
with
\begin{equation}
E_1(\kk) = \sqrt{[n^2(c_0+3c_1)^2+4n^2c_q(c_0+2c_1)]\epsk^2 - 4n^3c_1(c_0+3c_1)(c_1-c_q)\epsk+[2n^2c_1(c_1-c_q)]^2}.
\end{equation}
Defining
\begin{equation}
h_{00} \equiv  n(c_0+c_1-c_1),\quad h_{01} \equiv q\sin(2\theta)/2,\quad h_{11} \equiv -2nc_1+nc_q,\quad h_{211} \equiv nc_q
\end{equation}
and
\begin{align}
u_{\kk,\pm,1} = &\; -\big(h_{01}\big[\pm 2E_1+4\epsk^2+2\epsk(2E_{\kk,\pm}+h_{00}+2h_{11}-h_{211})\nonumber\\
&\;\qquad+(h_{11}-h_{211})(2E_{\kk,\pm}+h_{11}+h_{211})\big]\big)\big/\big(4\epsk h_{01}^2+2\epsk(h_{11}-h_{00})h_{211}\nonumber\\
&\qquad\qquad\qquad\qquad\qquad\qquad\qquad\qquad\qquad\qquad +h_{211}(\pm 2E_1+h_{11}^2-h_{211}^2)\big),\\
u_{\kk,\pm,2} =&\; \big(4\epsk^2(h_{00}-h_{11})-2(E_{\kk,\pm}+h_{11})(\pm 2E_1+h_{11}^2-h_{211}^2)\nonumber\\
&\;\qquad - 2\epsk(\pm 2E_1 + 4h_{01}^2-2E_{\kk,\pm}(h_{00}-h_{11})-2h_{00}h_{11}+3h_{11}^2\nonumber\\
&\;\qquad-h_{211}^2)\big)\big/\big(8\epsk h_{01}^2+4\epsk(h_{11}-h_{00})h_{211}+2h_{211}(\pm 2E_1+h_{11}^2- h_{211}^2)\big),\\
v_{\kk,\pm,1} =&\; \frac{h_{01}\big[\pm 2E_1+2\epsk(h_{00}+h_{211})+(h_{11}-h_{211})(2E_{\kk,\pm}+h_{11}+h_{211})\big]}{4\epsk h_{01}^2+2\epsk(h_{11}-h_{00})h_{211}+h_{211}(\pm 2E_1+h_{11}^2-h_{211}^2)},\\
v_{\kk,\pm,2} =&\; 1,
\end{align}
together with normalization factors
\begin{equation}
\mathcal{N}_{\kk,\pm} \equiv \sqrt{u_{\kk,\pm,1}^2+u_{\kk,\pm,2}^2-v_{\kk,\pm,1}^2-v_{\kk,\pm,2}^2}
\end{equation}
we find Bogoliubov transformation matrices (in the parametrization of \cite{uchino_bogoliubov_2010})
\begin{equation}
U_{\kk,d\theta} = \left(\begin{matrix}
u_{\kk,+,1}/\mathcal{N}_{\kk,+} & u_{\kk,+,2}/\mathcal{N}_{\kk,+} \\
u_{\kk,-,1}/\mathcal{N}_{\kk,-} & u_{\kk,-,2}/\mathcal{N}_{\kk,-}
\end{matrix}\right),\qquad V_{\kk, d\theta} = \left(\begin{matrix}
-v_{\kk,+,1}/\mathcal{N}_{\kk,+} & -v_{\kk,+,2}/\mathcal{N}_{\kk,+} \\
-v_{\kk,-,1}/\mathcal{N}_{\kk,-} & -v_{\kk,-,2}/\mathcal{N}_{\kk,-}
\end{matrix}\right).
\end{equation}
{These fulfil the identities}
\begin{equation}
U_{\kk,d\theta}U_{\kk,d\theta}^\dagger - V_{\kk,d\theta}V_{\kk,d\theta}^\dagger = 1, \qquad U_{\kk,d\theta}^*V_{\kk,d\theta}^\dagger - V_{\kk,d\theta}^*U_{\kk,d\theta}^\dagger = 0,\label{EqUVIdentities}
\end{equation}
as required for the transformations to preserve canonical commutation relations.
This requirement is not fulfilled for the transformations given in  \cite{uchino_bogoliubov_2010}.

The complete transformation matrices diagonalizing the Bogoliubov Hamiltonian (\ref{EqHamiltonianBogoliubov}) read
\begin{equation}
U_{\kk} = \left(\begin{matrix}
U_{\kk,d\theta} & 0\\
0 & u_{\kk,f_z}
\end{matrix}\right) A(\theta),\qquad V_{\kk} = \left(\begin{matrix}
V_{\kk,d\theta} & 0\\
0 & v_{\kk,f_z}
\end{matrix}\right) A(\theta).
\end{equation}

\subsection{Thermal structure factors from Bogoliubov theory}\label{SecSpectraFromBogTheory}
We provide analytical computations of thermal structure factors in the spinor Bose gas from Bogoliubov theory. We are interested in correlators of the form
\begin{equation}
\langle\hat{C}^\dagger(\xx)\hat{C}(\yy)\rangle_{\beta,s}\label{EqCorrelatorGeneral}
\end{equation}
for a composite field $\hat{C}(\xx)$ given by
\begin{equation}
\hat{C}(\xx) =\sum_{m,m'=-1}^{+1} \hat{\psi}_m^\dagger(\xx) c_{mm'} \hat{\psi}_{m'}(\xx)
\end{equation}
with $c_{mm'}$ a $3\times 3$ matrix corresponding to the type of spectrum under investigation; $c=f^x+if^y$ leads to the transversal magnetization spectrum, $c=f^z$ describes the spectrum of magnetization in $z$-direction, $c=\textrm{diag}(1,1,1)$ describes the total density spectrum. In Eq. (\ref{EqCorrelatorGeneral}) $\langle\cdot\rangle_{\beta,s}$ indicates the thermal expectation value at inverse temperature $\beta=1/(k_B T)$ with symmetrically (Weyl-) ordered arguments. We compare with symmetrically ordered predictions since expectation values of experimental observables are inferred from realizations of observables $O^{(i)}$ given by polynomials of complex numbers (cf. \cite{symes_static_2014} for a similar normal-ordered computation).
Fourier-transforming Eq. (\ref{EqCorrelatorGeneral}) with respect to the relative coordinate $\xx-\yy$, we obtain the structure factor
{
\begin{align}
\langle \hat{C}^\dagger(\kk)\hat{C}(\kk)\rangle_{\beta,s} =&\; \int d(\xx-\yy) \, \langle\hat{C}^\dagger(\xx)\hat{C}(\yy)\rangle_{\beta,s} \, e^{-i\kk (\xx-\yy)}\\
=&\; \frac{1}{V}\sum_{m,m',n,n'} c^\dagger_{mm'}c_{nn'} \sum_{\pp,\qq} \langle \hat{a}^\dagger_{\pp+\kk,m}\hat{a}_{\pp,m'}\hat{a}^\dagger_{\qq,n}\hat{a}_{\qq+\kk,n'}\rangle_{\beta,s}.\label{EqStructureFactorGeneral}
\end{align}
}
To total density structure factors a photon shot noise level of 0.6 after normalization is added.

In the Bogoliubov approximation Eq. (\ref{EqStructureFactorGeneral}) simplifies as follows. We replace zero modes of creation and annihilation operators by numbers, {$\hat{a}_{\mathbf{0},m} = \sqrt{N_{\mathrm{cond}}} \zeta_m$, $N_{\mathrm{cond}}$ the total number of condensate atoms}. Contributions from fluctuating modes $\hat{a}_{\kk\neq\mathbf{0},m}$ are computed via Wick's theorem. Symmetrically ordered propagators are defined as
\begin{align}
G^{11}_{mm'}(\kk) \equiv&\; \langle \hat{a}_{\kk,m}\hat{a}^\dagger_{\kk,m'}\rangle_{\beta,s},\qquad G^{22}_{mm'}(\kk) \equiv \langle \hat{a}^\dagger_{\kk,m}\hat{a}_{\kk,m'}\rangle_{\beta,s},\\
G^{12}_{mm'}(\kk) \equiv&\; \langle \hat{a}_{\kk,m}\hat{a}_{-\kk,m'}\rangle_{\beta,s},\qquad G^{21}_{mm'}(\kk) \equiv \langle \hat{a}^\dagger_{\kk,m}\hat{a}^\dagger_{-\kk,m'}\rangle_{\beta,s}.
\end{align}
The first two of these we refer to as normal propagators, the second two as anomalous propagators. Any normal propagator evaluated for non-diagonal momenta such as  $\langle \hat{a}^\dagger_{\kk,m}\hat{a}_{\kk',m'}\rangle_{\beta,s}$ for $\kk'\neq \kk$ and any anomalous propagator evaluated for non-anti-diagonal momenta such as $\langle \hat{a}_{\kk,m}\hat{a}_{-\kk',m'}\rangle_{\beta,s}$ for $\kk'\neq\kk$ equates to zero. We then find for $\kk = \mathbf{0}$,
{
\begin{align}
\langle \hat{C}^\dagger(\mathbf{0})\hat{C}(\mathbf{0})\rangle_{\beta,s} =&\; \frac{ N_{\mathrm{cond}}(N_{\mathrm{cond}}-1)}{V}\sum_{m,m',n,n'} c^\dagger_{mm'}c_{nn'}\zeta_m^*\zeta_n^*\zeta_{m'}\zeta_{n'} + \mathcal{O}(N_{\mathrm{cond}}),\label{EqStructureFactorZeroMode}
\end{align}
}
and for non-zero modes $\kk\neq\mathbf{0}$,
\begin{align}
\langle \hat{C}^\dagger(\mathbf{k})\hat{C}(\mathbf{k})\rangle_{\beta,s} =&\; \frac{N_{\mathrm{cond}}}{V}\sum_{m,m',n,n'} c^\dagger_{mm'}c_{nn'}\bigg[\zeta_m^*\zeta_n^* G^{12}_{m'n'}(\kk) +  \zeta_m^*\zeta_{n'}G^{11}_{m'n}(\kk)\nonumber\\
&\qquad\qquad\qquad + \zeta_n^*\zeta_{m'}G^{22}_{mn'}(\kk) + \zeta_{m'}\zeta_{n'}G^{21}_{mn}(\kk)\bigg] + \mathcal{O}(1),\label{EqStructureFactorNonZeroMode}
\end{align}
where we have used that the propagators $G^{ab}_{mm'}(\kk)$ only depend on the absolute value of the momentum, $|\kk|$. Experimental structure factors, normalized analogously, are compared to {$\langle \hat{C}^\dagger(\mathbf{k})\hat{C}(\mathbf{k})\rangle_{\beta,s}/n_{\mathrm{cond}}$, $n_{\mathrm{cond}}$ the total condensate atom density}. {The photon shot noise of the absorption imaging is determined to be $0.6$ after normalization and is added on the thermal prediction of the total density fluctuations.}

In the easy-plane ferromagnetic phase thermal propagators $G^{ab}_{mm'}(\kk\neq\mathbf{0})$ can be computed from the Bogoliubov transformations derived in Methods Sec. \ref{SecBogTrafosBAPhase}. Bogoliubov quasiparticle excitations {$f_z$ and $\pm$} are expressed in terms of fundamental magnetic sublevel excitations as
\begin{equation}
\left(\begin{matrix}
\hat{b}_{\kk,j}\\
\hat{b}^\dagger_{-\kk,j}
\end{matrix}\right) = \sum_{m=-1}^{+1} \left(\begin{matrix}
U_{\kk,jm} & V_{\kk,jm}\\
V_{-\kk,jm}^* & U_{-\kk,jm}^*
\end{matrix}\right) \left(\begin{matrix}
\hat{a}_{\kk,m}\\
\hat{a}_{-\kk,m}^\dagger
\end{matrix}\right),
\end{equation}
for $j\in\{f_z,\pm\}$. Inverting this, we obtain \cite{kawaguchi_spinor_2012}
\begin{equation}
\left(\begin{matrix}
\hat{a}_{\kk,m}\\
\hat{a}_{-\kk,m}^\dagger
\end{matrix}\right) = \sum_{j\in\{f_z,\pm\}}\left(\begin{matrix}
U_{\kk,mj}^\dagger & -V_{\kk,mj}^T\\
-V_{-\kk,mj}^\dagger & U_{-\kk,mj}^T
\end{matrix}\right)\left(\begin{matrix}
\hat{b}_{\kk,j}\\
\hat{b}^\dagger_{-\kk,j}
\end{matrix}\right).
\end{equation}
The propagators can now efficiently be computed from the tensor product
\begin{align}
&\left(\begin{matrix}
G^{12}_{mm'}(\kk) & G^{11}_{mm'}(\kk)\\
G^{22}_{mm'}(-\kk) & G^{21}_{mm'}(-\kk)
\end{matrix}\right) = \langle \left(\begin{matrix}
\hat{a}_{\kk,m}\\
\hat{a}_{-\kk,m}^\dagger
\end{matrix}\right)\otimes \left(\hat{a}_{-\kk,m'},\hat{a}^\dagger_{\kk,m'}\right)\rangle_{\beta,s}\\
&= \sum_{j,j'\in\{f_z,\pm\}}\left(\begin{matrix}
U_{\kk,mj}^\dagger & -V_{\kk,mj}^T\\
-V_{-\kk,mj}^\dagger & U_{-\kk,mj}^T
\end{matrix}\right)\left(\begin{matrix}
\langle \hat{b}_{\kk,j}\hat{b}_{-\kk,j'}\rangle_{\beta,s} & \langle \hat{b}_{\kk,j}\hat{b}^\dagger_{\kk,j'}\rangle_{\beta,s}\\
\langle \hat{b}^\dagger_{-\kk,j}\hat{b}_{-\kk,j'}\rangle_{\beta,s} & \langle \hat{b}^\dagger_{-\kk,j}\hat{b}^\dagger_{\kk,j'}\rangle_{\beta,s}
\end{matrix}\right)\left(\begin{matrix}
U_{-\kk,j'm'}^* & -V_{-\kk,j'm'}^*\\
-V_{\kk,j'm'} & U_{\kk,j'm'}
\end{matrix}\right).\label{EqPropagatorsFromBogoliubov}
\end{align}
The Bogoliubov quasiparticle modes $\hat{b}_{\kk,j}$ are occupied thermally,
\begin{equation}
\langle \hat{b}^\dagger_{\kk,j}\hat{b}_{\kk,j'}\rangle_\beta = \delta_{jj'}n_\beta(E_{\kk,j}),\qquad \langle \hat{b}_{\kk,j}\hat{b}^\dagger_{\kk,j'}\rangle_\beta = \delta_{jj'}(n_\beta(E_{\kk,j})+1),\label{EqOccbModes}
\end{equation}
with the Bose-Einstein distribution $n_\beta(E_{\kk,j}) \equiv 1/(\exp(\beta E_{\kk,j}) - 1)$. Anomalous propagators of $\hat{b}_{\kk,j}$-modes are zero. Insertion of (\ref{EqOccbModes}) into (\ref{EqPropagatorsFromBogoliubov}) and using that $U_{\kk,mj}$ and $V_{\kk,mj}$ only depend on $|\kk|$ leads to
\begin{align}
G^{11}_{mm'}(\kk) =&\; \sum_{j\in\{f_z,\pm\}} \bigg[ U^\dagger_{\kk,mj} \left(n_{\beta}(E_{\kk,j})+\frac{1}{2}\right) U_{\kk,jm'} + V^T_{\kk,mj}\left(n_{\beta}(E_{\kk,j})+\frac{1}{2}\right) V^*_{\kk,jm'}\bigg], \\
G^{22}_{mm'}(\kk) =&\; \sum_{j\in\{f_z,\pm\}}\bigg[ V^\dagger_{\kk,mj} \left(n_{\beta}(E_{\kk,j})+\frac{1}{2}\right) V_{\kk,jm'} + U^T_{\kk,mj}\left(n_{\beta}(E_{\kk,j})+\frac{1}{2}\right) U^*_{\kk,jm'}\bigg],\\
G^{12}_{mm'}(\kk) =&\; -\sum_{j\in\{f_z,\pm\}} \bigg[U^\dagger_{\kk,mj}\left(n_\beta(E_{\kk,j})+\frac{1}{2}\right) V_{\kk,jm'} + V^T_{\kk,mj}\left(n_{\beta}(E_{\kk,j})+\frac{1}{2}\right) U^*_{\kk,jm'}\bigg], \\
G^{21}_{mm'}(\kk) =&\; -\sum_{j\in\{f_z,\pm\}} \bigg[V^\dagger_{\kk,mj}\left(n_\beta(E_{\kk,j})+\frac{1}{2}\right) U_{\kk,jm'} + U^T_{\kk,mj}\left(n_{\beta}(E_{\kk,j})+\frac{1}{2}\right)V^*_{\kk,jm'}\bigg].
\end{align}
With these expressions thermal  structure factors can be readily computed from Eq. (\ref{EqStructureFactorNonZeroMode}).

\subsection{Fitting thermal Bogoliubov theory structure factors}
Using a least-squares fitting procedure and Gibbs sampling, systematic as well as statistical uncertainties on the optimal set of parameters are estimated. Given experimental  structure factors $S_{\hat{C},\exp}(k)=\langle \hat{C}^\dagger(k)\hat{C}(k)\rangle$ with $\hat{C}\in \mathcal{S}:=\{\hat{N}_{+1},\hat{N}_0,\hat{N}_{-1},\hat{F}_z,\hat{F}_\perp\}$, we determine an optimal set of parameters $T,q,nc_1$ by minimizing
\begin{equation}\label{EqDefChiSquare}
\chi^2(T,q,nc_1;k_{\max}) = \sum_{\hat{C}\in\mathcal{S}}\sum_{k}^{k_{\max}} \frac{(S_{\hat{C},\exp}(k) - S_{\hat{C},\Bog}(k;T,q,nc_1))^2}{\Delta S_{\hat{C},\exp}(k)^2},
\end{equation}
with $S_{\hat{C},\Bog}(k;T,q,nc_1) = \langle \hat{C}^\dagger(k)\hat{C}(k)\rangle_{1/(k_B T),s}$ the Bogoliubov theory  structure factor computed for parameters $T,q,nc_1$, and $\Delta S_{\hat{C},\exp}(k)$ the standard deviation of $S_{\hat{C},\exp}(k)$ computed from experimental realizations. {Technical correlations of the absorption imaging are described by real-space signals convoluted with a Gaussian of r.\,m.\,s.\,width $w = 5.0\,\um$, taken into account by the multiplication of momentum-space structure factors with a Gaussian of width $2\pi/w$ \cite{kunkel_spatially_2018}.}
 Throughout the fitting procedure we set $c_0/c_1 \simeq -216$ in accordance with \cite{stamper-kurn_spinor_2013}. In the definition of $\chi^2$ we did not include {the structure factor of the total density}.

We define a {distribution} of parameters for specific $k_{\max}$,
\begin{equation}
W(T,q,nc_1;k_{\max}) \sim \exp \left(-\chi^2(T,q,nc_1;k_{\max})/2\right).
\end{equation} 
We exploit Gibbs sampling to draw $i=1,\dots,100$ approximately i.i.d. samples $(T^{(i)}(k_{\max})$, $q^{(i)}(k_{\max})$, $nc_1^{(i)}(k_{\max}))$ from $W(T,q,nc_1;k_{\max})${, which only requires corresponding conditional distributions normalized individually}. For each $k_{\max}$ we compute their mean $(\Tb(k_{\max}),\qb(k_{\max}),\nconeb(k_{\max}))$. We repeat this for 5 values of $k_{\max}$ evenly spaced between {$0.1\cdot 2\pi/\um$ and $0.2\cdot 2\pi/\um$}. 
Collecting all $5\cdot 100$ samples in a single array, we take the mean values $\Tb,\qb,\nconeb$ of all samples as final parameter estimates and their distances to the boundaries of 68\% confidence intervals as corresponding error estimates.
Errors include systematic fit uncertainties. We obtain the final fit parameters
\begin{equation}
T = (57.4\pm 2.9)\,\mathrm{Hz},\quad q =  (0.30\pm 0.08)\,\mathrm{Hz}, \quad nc_1 = (-1.17\pm 0.25)\,\mathrm{Hz},
\end{equation}
{such that $nc_0 = (252\pm 54)\,\mathrm{Hz}$ and $q/(nc_1) = (-0.26\pm 0.09)$.}

\subsection{Drawing Bogoliubov theory samples}
Bogoliubov theory being quadratic in fluctuating field creation and annihilation operators, it is fully described by zero modes and a suitable covariance matrix of fluctuations. The latter can be constructed from the propagators $G^{ij}_{mm'}(\kk)$.

Given a one-dimensional real-space lattice {$\{-\mathcal{N},\dots, \mathcal{N}\}\cdot a$} with lattice spacing {$a = L/(2\mathcal{N})$}, the corresponding momentum-space lattice reads $\{-\mathcal{N},\dots,\mathcal{N}\}\cdot \pi/(\mathcal{N} a)$. With $\Delta z = z_1-z_2 \in \{-2\mathcal{N},\dots,2\mathcal{N}\}$ for $z_i \in \{-\mathcal{N},\dots, \mathcal{N}\}$ we compute real-space propagators via
\begin{equation}
{\tilde{G}^{ij}_{mm'}(\Delta z) = \frac{1}{L}\sum_{p=-\mathcal{N}}^{\mathcal{N}} G^{ij}_{mm'}(p)\exp \left(\frac{2\pi i p \Delta z}{2\mathcal{N}+1}\right).}
\end{equation}
We assemble these into magnetic sublevel-specific covariance matrices,
{
\tiny
\begin{align}
&\;\textrm{Cov}_{mm'} = \langle \left(\begin{matrix}
\hat{\psi}_m(-\mathcal{N})\\
\vdots\\
\hat{\psi}_m(\mathcal{N})\\
\hat{\psi}_m^\dagger(-\mathcal{N})\\
\vdots\\
\hat{\psi}_m^\dagger(\mathcal{N})
\end{matrix}\right) \left(\begin{matrix}
\hat{\psi}_{m'}(-\mathcal{N}), \dots, \hat{\psi}_{m'}(\mathcal{N}),\hat{\psi}^\dagger_{m'}(-\mathcal{N}),\dots, \hat{\psi}^\dagger_{m'}(\mathcal{N})
\end{matrix}\right)\rangle_{\beta,s}\\
&\;= \left(\begin{matrix}
\tilde{G}^{12}_{mm'}(0) & \tilde{G}^{12}_{mm'}(-1) & \dots & \tilde{G}^{12}_{mm'}(-2\mathcal{N}) & \tilde{G}^{11}_{mm'}(0) & \tilde{G}^{11}_{mm'}(-1) & \dots & \tilde{G}^{11}_{mm'}(-2\mathcal{N}) \\
\tilde{G}^{12}_{mm'}(1) & \tilde{G}^{12}_{mm'}(0) & \dots & \tilde{G}^{12}_{mm'}(-2\mathcal{N}+1) & \tilde{G}^{11}_{mm'}(1) & \tilde{G}^{11}_{mm'}(0) & \dots & \tilde{G}^{11}_{mm'}(-2\mathcal{N}+1) \\
\vdots & \vdots & \ddots & \vdots & \vdots & \vdots & \ddots & \vdots\\
\tilde{G}^{12}_{mm'}(2\mathcal{N}) & \tilde{G}^{12}_{mm'}(2\mathcal{N}-1) & \dots & \tilde{G}^{12}_{mm'}(0) & \tilde{G}^{11}_{mm'}(2\mathcal{N}) & \tilde{G}^{11}_{mm'}(2\mathcal{N}-1) & \dots & \tilde{G}^{11}_{mm'}(0)\\
\tilde{G}^{22}_{mm'}(0) & \tilde{G}^{22}_{mm'}(-1) & \dots & \tilde{G}^{22}_{mm'}(-2\mathcal{N}) & \tilde{G}^{21}_{mm'}(0) & \tilde{G}^{21}_{mm'}(-1) & \dots & \tilde{G}^{21}_{mm'}(-2\mathcal{N}) \\
\tilde{G}^{22}_{mm'}(1) & \tilde{G}^{22}_{mm'}(0) & \dots & \tilde{G}^{22}_{mm'}(-2\mathcal{N}+1) & \tilde{G}^{21}_{mm'}(1) & \tilde{G}^{21}_{mm'}(0) & \dots & \tilde{G}^{21}_{mm'}(-2\mathcal{N}+1) \\
\vdots & \vdots & \ddots & \vdots & \vdots & \vdots & \ddots & \vdots\\
\tilde{G}^{22}_{mm'}(2\mathcal{N}) & \tilde{G}^{22}_{mm'}(2\mathcal{N}-1) & \dots & \tilde{G}^{22}_{mm'}(0) & \tilde{G}^{21}_{mm'}(2\mathcal{N}) & \tilde{G}^{21}_{mm'}(2\mathcal{N}-1) & \dots & \tilde{G}^{21}_{mm'}(0)\end{matrix}\right),
\end{align}
}having exploited spatial homogeneity. We decompose complex field operators into real components, $\hat{\psi}(x) = \frac{1}{\sqrt{2}} ( \hat{\psi}_1(x) + i \hat{\psi}_2(x))$, translating into the unitary transformation
\begin{equation}
\left(\begin{matrix}
\hat{\psi}_{m,1}(-\mathcal{N})\\
\vdots\\
\hat{\psi}_{m,1}(\mathcal{N})\\
\hat{\psi}_{m,2}(-\mathcal{N})\\
\vdots\\
\hat{\psi}_{m,2}(\mathcal{N})
\end{matrix}\right) = A\left(\begin{matrix}
\hat{\psi}_m(-\mathcal{N})\\
\vdots\\
\hat{\psi}_m(\mathcal{N})\\
\hat{\psi}_m^\dagger(-\mathcal{N})\\
\vdots\\
\hat{\psi}_m^\dagger(\mathcal{N})
\end{matrix}\right),\qquad A = \frac{1}{\sqrt{2}}\left(\begin{matrix}
I & I\\
-i I & i I
\end{matrix}\right),
\end{equation}
with $I$ the $(2\mathcal{N}+1)\times(2\mathcal{N}+1)$-dimensional identity matrix. We define the final covariance matrix of the theory as
\begin{equation}
\textrm{Cov} = \left(\begin{matrix}
A\, \textrm{Cov}_{+1,+1} A^T & A\, \textrm{Cov}_{+1,0} A^T & A \,\textrm{Cov}_{+1,-1} A^T\\
A\, \textrm{Cov}_{0,+1} A^T & A \,\textrm{Cov}_{0,0} A^T & A\, \textrm{Cov}_{0,-1} A^T
\\
A\, \textrm{Cov}_{-1,+1} A^T & A\, \textrm{Cov}_{-1,0} A^T & A \,\textrm{Cov}_{-1,-1} A^T\end{matrix}\right).
\end{equation}
Finally, we sample $i=1,\dots, N_{\mathrm{sample}}$ field realizations
\begin{equation}
\psi^{(i)}  = \left(\begin{matrix}
\psi^{(i)}_{+1}\\
\psi^{(i)}_0\\
\psi^{(i)}_{-1}
\end{matrix}\right), \qquad \psi^{(i)}_m = \left(\begin{matrix}
\psi^{(i)}_{m,1}(-\mathcal{N}), & \cdots, & \psi^{(i)}_{m,1}(\mathcal{N}), & \psi^{(i)}_{m,2}(-\mathcal{N}),&\cdots, & \psi^{(i)}_{m,2}(\mathcal{N})
\end{matrix}\right)^T,
\end{equation}
from the multivariate Gaussian distribution with zero mean vector and covariance matrix $\mathrm{Cov}$. This corresponds to samples from the Wigner distribution of the symmetrically ordered Bogoliubov theory of fluctuating modes at inverse temperature $\beta$. We sample fields in position space instead of momentum space, since in momentum space, having decomposed the operators $\hat{a}_{k,m}$ into $\hat{a}_{k,m} = (\hat{a}_{k,m,1}+i\hat{a}_{k,m,2})/\sqrt{2}$, the components $\hat{a}_{k,m,j}$ need to satisfy $\hat{a}_{k,m,j}^\dagger = \hat{a}_{-k,m,j}$, such that samples of individual momentum modes cannot be drawn independently.
From the fluctuating realizations $\psi^{(i)}$ we can compute realizations of the individual spin sublevel fields in real space,
\begin{equation}\label{EqBogoliubovShotFinalComp}
\psi^{(i)}_m(x) = \frac{1}{\sqrt{2}}[\psi^{(i)}_{m,1}(x) + i \psi^{(i)}_{m,2}(x)] + \sqrt{n_{\mathrm{cond}}}\zeta_m,
\end{equation}
with $n_{\mathrm{cond}}$ the condensate density. We explicitely checked that for increasing sample numbers structure factors computed from samples $\psi^{(i)}_m(x)$ converge towards their expectations $G^{ij}_{mm'}(k)$.

The composite operator histograms displayed in Ext. Data Fig. 2 are computed from composite profiles of individual realizations given by {$\sum_{m,m'=-1}^{+1}(\psi^{(i)}_m(x))^* c_{mm'} \psi^{(i)}_{m'}(x)/\sqrt{n_{\mathrm{cond}}}$} with matrices $c$ as denoted in Methods \ref{SecSpectraFromBogTheory}.
 
\  

\,
\clearpage
\bibliographystyle{PrueferThermalSpin}
\bibliography{PrueferThermalSpin}

\begin{thebibliography}{10}
\expandafter\ifx\csname url\endcsname\relax
  \def\url#1{\texttt{#1}}\fi
\expandafter\ifx\csname urlprefix\endcsname\relax\def\urlprefix{URL }\fi
\providecommand{\bibinfo}[2]{#2}
\providecommand{\eprint}[2][]{\url{#2}}

\bibitem{zutic_spintronics_2004}
\bibinfo{author}{Žutić, I.}, \bibinfo{author}{Fabian, J.} \&
  \bibinfo{author}{Das~Sarma, S.}
\newblock
  \href{http://dx.doi.org/10.1103/RevModPhys.76.323}{\bibinfo{title}{Spintronics:
  {Fundamentals} and applications}}.
\newblock \emph{\bibinfo{journal}{Reviews of Modern Physics}}
  \textbf{\bibinfo{volume}{76}}, \bibinfo{pages}{323--410}
  (\bibinfo{year}{2004}).

\bibitem{sonin_spin_2010}
\bibinfo{author}{Sonin, E.~B.}
\newblock
  \href{http://dx.doi.org/10.1080/00018731003739943}{\bibinfo{title}{Spin
  currents and spin superfluidity}}.
\newblock \emph{\bibinfo{journal}{Advances in Physics}}
  \textbf{\bibinfo{volume}{59}}, \bibinfo{pages}{181--255}
  (\bibinfo{year}{2010}).

\bibitem{stamper-kurn_spinor_2013}
\bibinfo{author}{Stamper-Kurn, D.~M.} \& \bibinfo{author}{Ueda, M.}
\newblock
  \href{http://dx.doi.org/10.1103/RevModPhys.85.1191}{\bibinfo{title}{Spinor
  {Bose} gases: {Symmetries}, magnetism, and quantum dynamics}}.
\newblock \emph{\bibinfo{journal}{Reviews of Modern Physics}}
  \textbf{\bibinfo{volume}{85}}, \bibinfo{pages}{1191--1244}
  (\bibinfo{year}{2013}).

\bibitem{sadler_spontaneous_2006}
\bibinfo{author}{Sadler, L.~E.}, \bibinfo{author}{Higbie, J.~M.},
  \bibinfo{author}{Leslie, S.~R.}, \bibinfo{author}{Vengalattore, M.} \&
  \bibinfo{author}{Stamper-Kurn, D.~M.}
\newblock
  \href{http://dx.doi.org/10.1038/nature05094}{\bibinfo{title}{Spontaneous
  symmetry breaking in a quenched ferromagnetic spinor {Bose}–{Einstein}
  condensate}}.
\newblock \emph{\bibinfo{journal}{Nature}} \textbf{\bibinfo{volume}{443}},
  \bibinfo{pages}{312--315} (\bibinfo{year}{2006}).

\bibitem{kunkel_simultaneous_2019}
\bibinfo{author}{Kunkel, P.} \emph{et~al.}
\newblock
  \href{http://dx.doi.org/10.1103/PhysRevLett.123.063603}{\bibinfo{title}{Simultaneous
  {Readout} of {Noncommuting} {Collective} {Spin} {Observables} beyond the
  {Standard} {Quantum} {Limit}}}.
\newblock \emph{\bibinfo{journal}{Physical Review Letters}}
  \textbf{\bibinfo{volume}{123}}, \bibinfo{pages}{063603}
  (\bibinfo{year}{2019}).

\bibitem{glauber_quantum_1963}
\bibinfo{author}{Glauber, R.~J.}
\newblock \href{http://dx.doi.org/10.1103/PhysRev.130.2529}{\bibinfo{title}{The
  {Quantum} {Theory} of {Optical} {Coherence}}}.
\newblock \emph{\bibinfo{journal}{Physical Review}}
  \textbf{\bibinfo{volume}{130}}, \bibinfo{pages}{2529--2539}
  (\bibinfo{year}{1963}).

\bibitem{landau_theory_1941}
\bibinfo{author}{Landau, L.}
\newblock
  \href{http://dx.doi.org/10.1103/PhysRev.60.356}{\bibinfo{title}{Theory of the
  {Superfluidity} of {Helium} {II}}}.
\newblock \emph{\bibinfo{journal}{Physical Review}}
  \textbf{\bibinfo{volume}{60}}, \bibinfo{pages}{356--358}
  (\bibinfo{year}{1941}).

\bibitem{uchino_bogoliubov_2010}
\bibinfo{author}{Uchino, S.}, \bibinfo{author}{Kobayashi, M.} \&
  \bibinfo{author}{Ueda, M.}
\newblock
  \href{http://dx.doi.org/10.1103/PhysRevA.81.063632}{\bibinfo{title}{Bogoliubov
  theory and {Lee}-{Huang}-{Yang} corrections in spin-1 and spin-2
  {Bose}-{Einstein} condensates in the presence of the quadratic {Zeeman}
  effect}}.
\newblock \emph{\bibinfo{journal}{Physical Review A}}
  \textbf{\bibinfo{volume}{81}}, \bibinfo{pages}{063632}
  (\bibinfo{year}{2010}).

\bibitem{phuc_effects_2011}
\bibinfo{author}{Phuc, N.~T.}, \bibinfo{author}{Kawaguchi, Y.} \&
  \bibinfo{author}{Ueda, M.}
\newblock
  \href{http://dx.doi.org/10.1103/PhysRevA.84.043645}{\bibinfo{title}{Effects
  of thermal and quantum fluctuations on the phase diagram of a spin-1 87 {Rb}
  {Bose}-{Einstein} condensate}}.
\newblock \emph{\bibinfo{journal}{Physical Review A}}
  \textbf{\bibinfo{volume}{84}}, \bibinfo{pages}{043645}
  (\bibinfo{year}{2011}).

\bibitem{symes_static_2014}
\bibinfo{author}{Symes, L.~M.}, \bibinfo{author}{Baillie, D.} \&
  \bibinfo{author}{Blakie, P.~B.}
\newblock
  \href{http://dx.doi.org/10.1103/PhysRevA.89.053628}{\bibinfo{title}{Static
  structure factors for a spin-1 {Bose}-{Einstein} condensate}}.
\newblock \emph{\bibinfo{journal}{Physical Review A}}
  \textbf{\bibinfo{volume}{89}}, \bibinfo{pages}{053628}
  (\bibinfo{year}{2014}).

\bibitem{bloch_quantum_2012}
\bibinfo{author}{Bloch, I.}, \bibinfo{author}{Dalibard, J.} \&
  \bibinfo{author}{Nascimbène, S.}
\newblock \href{http://dx.doi.org/10.1038/nphys2259}{\bibinfo{title}{Quantum
  simulations with ultracold quantum gases}}.
\newblock \emph{\bibinfo{journal}{Nature Physics}}
  \textbf{\bibinfo{volume}{8}}, \bibinfo{pages}{267--276}
  (\bibinfo{year}{2012}).

\bibitem{gross_quantum_2017}
\bibinfo{author}{Gross, C.} \& \bibinfo{author}{Bloch, I.}
\newblock
  \href{http://dx.doi.org/10.1126/science.aal3837}{\bibinfo{title}{Quantum
  simulations with ultracold atoms in optical lattices}}.
\newblock \emph{\bibinfo{journal}{Science}} \textbf{\bibinfo{volume}{357}},
  \bibinfo{pages}{995--1001} (\bibinfo{year}{2017}).

\bibitem{rigol_thermalization_2008}
\bibinfo{author}{Rigol, M.}, \bibinfo{author}{Dunjko, V.} \&
  \bibinfo{author}{Olshanii, M.}
\newblock
  \href{http://dx.doi.org/10.1038/nature06838}{\bibinfo{title}{Thermalization
  and its mechanism for generic isolated quantum systems}}.
\newblock \emph{\bibinfo{journal}{Nature}} \textbf{\bibinfo{volume}{452}},
  \bibinfo{pages}{854--858} (\bibinfo{year}{2008}).

\bibitem{eisert_quantum_2015}
\bibinfo{author}{Eisert, J.}, \bibinfo{author}{Friesdorf, M.} \&
  \bibinfo{author}{Gogolin, C.}
\newblock \href{http://dx.doi.org/10.1038/nphys3215}{\bibinfo{title}{Quantum
  many-body systems out of equilibrium}}.
\newblock \emph{\bibinfo{journal}{Nature Physics}}
  \textbf{\bibinfo{volume}{11}}, \bibinfo{pages}{124--130}
  (\bibinfo{year}{2015}).

\bibitem{reimann_typical_2016}
\bibinfo{author}{Reimann, P.}
\newblock \href{http://dx.doi.org/10.1038/ncomms10821}{\bibinfo{title}{Typical
  fast thermalization processes in closed many-body systems}}.
\newblock \emph{\bibinfo{journal}{Nature Communications}}
  \textbf{\bibinfo{volume}{7}}, \bibinfo{pages}{10821} (\bibinfo{year}{2016}).

\bibitem{gogolin_equilibration_2016}
\bibinfo{author}{Gogolin, C.} \& \bibinfo{author}{Eisert, J.}
\newblock
  \href{http://dx.doi.org/10.1088/0034-4885/79/5/056001}{\bibinfo{title}{Equilibration,
  thermalisation, and the emergence of statistical mechanics in closed quantum
  systems}}.
\newblock \emph{\bibinfo{journal}{Reports on Progress in Physics}}
  \textbf{\bibinfo{volume}{79}}, \bibinfo{pages}{056001}
  (\bibinfo{year}{2016}).

\bibitem{mori_thermalization_2018}
\bibinfo{author}{Mori, T.}, \bibinfo{author}{Ikeda, T.~N.},
  \bibinfo{author}{Kaminishi, E.} \& \bibinfo{author}{Ueda, M.}
\newblock
  \href{http://dx.doi.org/10.1088/1361-6455/aabcdf}{\bibinfo{title}{Thermalization
  and prethermalization in isolated quantum systems: a theoretical overview}}.
\newblock \emph{\bibinfo{journal}{Journal of Physics B: Atomic, Molecular and
  Optical Physics}} \textbf{\bibinfo{volume}{51}}, \bibinfo{pages}{112001}
  (\bibinfo{year}{2018}).

\bibitem{kaufman_quantum_2016}
\bibinfo{author}{Kaufman, A.~M.} \emph{et~al.}
\newblock
  \href{http://dx.doi.org/10.1126/science.aaf6725}{\bibinfo{title}{Quantum
  thermalization through entanglement in an isolated many-body system}}.
\newblock \emph{\bibinfo{journal}{Science}} \textbf{\bibinfo{volume}{353}},
  \bibinfo{pages}{794--800} (\bibinfo{year}{2016}).

\bibitem{neill_ergodic_2016}
\bibinfo{author}{Neill, C.} \emph{et~al.}
\newblock \href{http://dx.doi.org/10.1038/nphys3830}{\bibinfo{title}{Ergodic
  dynamics and thermalization in an isolated quantum system}}.
\newblock \emph{\bibinfo{journal}{Nature Physics}}
  \textbf{\bibinfo{volume}{12}}, \bibinfo{pages}{1037--1041}
  (\bibinfo{year}{2016}).

\bibitem{clos_time-resolved_2016}
\bibinfo{author}{Clos, G.}, \bibinfo{author}{Porras, D.},
  \bibinfo{author}{Warring, U.} \& \bibinfo{author}{Schaetz, T.}
\newblock
  \href{http://dx.doi.org/10.1103/PhysRevLett.117.170401}{\bibinfo{title}{Time-{Resolved}
  {Observation} of {Thermalization} in an {Isolated} {Quantum} {System}}}.
\newblock \emph{\bibinfo{journal}{Physical Review Letters}}
  \textbf{\bibinfo{volume}{117}}, \bibinfo{pages}{170401}
  (\bibinfo{year}{2016}).

\bibitem{evrard_many-body_2021}
\bibinfo{author}{Evrard, B.}, \bibinfo{author}{Qu, A.},
  \bibinfo{author}{Dalibard, J.} \& \bibinfo{author}{Gerbier, F.}
\newblock
  \href{http://dx.doi.org/10.1103/PhysRevLett.126.063401}{\bibinfo{title}{From
  many-body oscillations to thermalization in an isolated spinor gas}}.
\newblock \emph{\bibinfo{journal}{Physical Review Letters}}
  \textbf{\bibinfo{volume}{126}}, \bibinfo{pages}{063401}
  (\bibinfo{year}{2021}).

\bibitem{griffin_bose-einstein_1995}
\bibinfo{editor}{Griffin, A.}, \bibinfo{editor}{Snoke, D.~W.} \&
  \bibinfo{editor}{Stringari, S.} (eds.) \emph{\bibinfo{title}{Bose-{Einstein}
  {Condensation}}} (\bibinfo{publisher}{Cambridge University Press},
  \bibinfo{year}{1995}).

\bibitem{bloch_measurement_2000}
\bibinfo{author}{Bloch, I.}, \bibinfo{author}{Hänsch, T.~W.} \&
  \bibinfo{author}{Esslinger, T.}
\newblock \href{http://dx.doi.org/10.1038/35003132}{\bibinfo{title}{Measurement
  of the spatial coherence of a trapped {Bose} gas at the phase transition}}.
\newblock \emph{\bibinfo{journal}{Nature}} \textbf{\bibinfo{volume}{403}},
  \bibinfo{pages}{166--170} (\bibinfo{year}{2000}).

\bibitem{deng_spatial_2007}
\bibinfo{author}{Deng, H.}, \bibinfo{author}{Solomon, G.~S.},
  \bibinfo{author}{Hey, R.}, \bibinfo{author}{Ploog, K.~H.} \&
  \bibinfo{author}{Yamamoto, Y.}
\newblock
  \href{http://dx.doi.org/10.1103/PhysRevLett.99.126403}{\bibinfo{title}{Spatial
  {Coherence} of a {Polariton} {Condensate}}}.
\newblock \emph{\bibinfo{journal}{Physical Review Letters}}
  \textbf{\bibinfo{volume}{99}}, \bibinfo{pages}{126403}
  (\bibinfo{year}{2007}).

\bibitem{damm_first-order_2017}
\bibinfo{author}{Damm, T.}, \bibinfo{author}{Dung, D.},
  \bibinfo{author}{Vewinger, F.}, \bibinfo{author}{Weitz, M.} \&
  \bibinfo{author}{Schmitt, J.}
\newblock
  \href{http://dx.doi.org/10.1038/s41467-017-00270-8}{\bibinfo{title}{First-order
  spatial coherence measurements in a thermalized two-dimensional photonic
  quantum gas}}.
\newblock \emph{\bibinfo{journal}{Nature Communications}}
  \textbf{\bibinfo{volume}{8}}, \bibinfo{pages}{158} (\bibinfo{year}{2017}).

\bibitem{proukakis_bose-einstein_2017-2}
\bibinfo{author}{Kollath, C.}, \bibinfo{author}{Giamarchi, T.} \&
  \bibinfo{author}{Rüegg, C.}
\newblock \bibinfo{title}{Bose-{Einstein} {Condensation} in {Quantum}
  {Magnets}}.
\newblock In \bibinfo{editor}{Proukakis, N.~P.}, \bibinfo{editor}{Snoke, D.~W.}
  \& \bibinfo{editor}{Littlewood, P.~B.} (eds.)
  \emph{\bibinfo{booktitle}{{Universal Themes of Bose-Einstein Condensation}}},
  \bibinfo{pages}{549--568} (\bibinfo{publisher}{Cambridge University Press},
  \bibinfo{year}{2017}).

\bibitem{jepsen_spin_2020}
\bibinfo{author}{Jepsen, P.~N.} \emph{et~al.}
\newblock
  \href{http://dx.doi.org/10.1038/s41586-020-3033-y}{\bibinfo{title}{Spin
  transport in a tunable {Heisenberg} model realized with ultracold atoms}}.
\newblock \emph{\bibinfo{journal}{Nature}} \textbf{\bibinfo{volume}{588}},
  \bibinfo{pages}{403--407} (\bibinfo{year}{2020}).

\bibitem{geier2021floquet}
\bibinfo{author}{Geier, S.} \emph{et~al.}
\newblock
  \href{http://dx.doi.org/10.1126/science.abd9547}{\bibinfo{title}{Floquet
  hamiltonian engineering of an isolated many-body spin system}}.
\newblock \emph{\bibinfo{journal}{Science}} \textbf{\bibinfo{volume}{374}},
  \bibinfo{pages}{1149--1152} (\bibinfo{year}{2021}).

\bibitem{scholl_microwave-engineering_2022}
\bibinfo{author}{Scholl, P.} \emph{et~al.}
\newblock \bibinfo{title}{Microwave-engineering of programmable {XXZ}
  {Hamiltonians} in arrays of {Rydberg} atoms}.
\newblock \emph{\bibinfo{journal}{arXiv:2107.14459}}  (\bibinfo{year}{2022}).

\bibitem{uchino_spinor_2015}
\bibinfo{author}{Uchino, S.}
\newblock
  \href{http://dx.doi.org/10.1103/PhysRevA.91.033605}{\bibinfo{title}{Spinor
  {Bose} gas in an elongated trap}}.
\newblock \emph{\bibinfo{journal}{Physical Review A}}
  \textbf{\bibinfo{volume}{91}}, \bibinfo{pages}{033605}
  (\bibinfo{year}{2015}).

\bibitem{navon_quantum_2021}
\bibinfo{author}{Navon, N.}, \bibinfo{author}{Smith, R.~P.} \&
  \bibinfo{author}{Hadzibabic, Z.}
\newblock
  \href{http://dx.doi.org/10.1038/s41567-021-01403-z}{\bibinfo{title}{Quantum
  gases in optical boxes}}.
\newblock \emph{\bibinfo{journal}{Nature Physics}}
  \textbf{\bibinfo{volume}{17}}, \bibinfo{pages}{1334--1341}
  (\bibinfo{year}{2021}).

\bibitem{gerbier_resonant_2006}
\bibinfo{author}{Gerbier, F.}, \bibinfo{author}{Widera, A.},
  \bibinfo{author}{Fölling, S.}, \bibinfo{author}{Mandel, O.} \&
  \bibinfo{author}{Bloch, I.}
\newblock
  \href{http://dx.doi.org/10.1103/PhysRevA.73.041602}{\bibinfo{title}{Resonant
  control of spin dynamics in ultracold quantum gases by microwave dressing}}.
\newblock \emph{\bibinfo{journal}{Physical Review A}}
  \textbf{\bibinfo{volume}{73}}, \bibinfo{pages}{041602}
  (\bibinfo{year}{2006}).

\bibitem{prufer_experimental_2020}
\bibinfo{author}{Prüfer, M.} \emph{et~al.}
\newblock
  \href{http://dx.doi.org/10.1038/s41567-020-0933-6}{\bibinfo{title}{Experimental
  extraction of the quantum effective action for a non-equilibrium many-body
  system}}.
\newblock \emph{\bibinfo{journal}{Nature Physics}}
  \textbf{\bibinfo{volume}{16}}, \bibinfo{pages}{1012--1016}
  (\bibinfo{year}{2020}).

\bibitem{kawaguchi_spinor_2012}
\bibinfo{author}{Kawaguchi, Y.} \& \bibinfo{author}{Ueda, M.}
\newblock
  \href{http://dx.doi.org/10.1016/j.physrep.2012.07.005}{\bibinfo{title}{Spinor
  {Bose}-{Einstein} condensates}}.
\newblock \emph{\bibinfo{journal}{Physics Reports}}
  \textbf{\bibinfo{volume}{520}}, \bibinfo{pages}{253--381}
  (\bibinfo{year}{2012}).

\bibitem{kohl_growth_2002}
\bibinfo{author}{Köhl, M.}, \bibinfo{author}{Davis, M.~J.},
  \bibinfo{author}{Gardiner, C.~W.}, \bibinfo{author}{Hänsch, T.~W.} \&
  \bibinfo{author}{Esslinger, T.}
\newblock
  \href{http://dx.doi.org/10.1103/PhysRevLett.88.080402}{\bibinfo{title}{Growth
  of {Bose}-{Einstein} {Condensates} from {Thermal} {Vapor}}}.
\newblock \emph{\bibinfo{journal}{Physical Review Letters}}
  \textbf{\bibinfo{volume}{88}}, \bibinfo{pages}{080402}
  (\bibinfo{year}{2002}).

\bibitem{ritter_observing_2007}
\bibinfo{author}{Ritter, S.} \emph{et~al.}
\newblock
  \href{http://dx.doi.org/10.1103/PhysRevLett.98.090402}{\bibinfo{title}{Observing
  the {Formation} of {Long}-{Range} {Order} during {Bose}-{Einstein}
  {Condensation}}}.
\newblock \emph{\bibinfo{journal}{Physical Review Letters}}
  \textbf{\bibinfo{volume}{98}}, \bibinfo{pages}{090402}
  (\bibinfo{year}{2007}).

\bibitem{kunkel_detecting_2022}
\bibinfo{author}{Kunkel, P.} \emph{et~al.}
\newblock
  \href{http://dx.doi.org/10.1103/PhysRevLett.128.020402}{\bibinfo{title}{Detecting
  {Entanglement} {Structure} in {Continuous} {Many}-{Body} {Quantum}
  {Systems}}}.
\newblock \emph{\bibinfo{journal}{Physical Review Letters}}
  \textbf{\bibinfo{volume}{128}}, \bibinfo{pages}{020402}
  (\bibinfo{year}{2022}).

\bibitem{naraschewski_spatial_1999}
\bibinfo{author}{Naraschewski, M.} \& \bibinfo{author}{Glauber, R.~J.}
\newblock
  \href{http://dx.doi.org/10.1103/PhysRevA.59.4595}{\bibinfo{title}{Spatial
  coherence and density correlations of trapped {Bose} gases}}.
\newblock \emph{\bibinfo{journal}{Physical Review A}}
  \textbf{\bibinfo{volume}{59}}, \bibinfo{pages}{4595--4607}
  (\bibinfo{year}{1999}).

\bibitem{ottl_correlations_2005}
\bibinfo{author}{Öttl, A.}, \bibinfo{author}{Ritter, S.},
  \bibinfo{author}{Köhl, M.} \& \bibinfo{author}{Esslinger, T.}
\newblock
  \href{http://dx.doi.org/10.1103/PhysRevLett.95.090404}{\bibinfo{title}{Correlations
  and {Counting} {Statistics} of an {Atom} {Laser}}}.
\newblock \emph{\bibinfo{journal}{Physical Review Letters}}
  \textbf{\bibinfo{volume}{95}}, \bibinfo{pages}{090404}
  (\bibinfo{year}{2005}).

\bibitem{hodgman_direct_2011}
\bibinfo{author}{Hodgman, S.~S.}, \bibinfo{author}{Dall, R.~G.},
  \bibinfo{author}{Manning, A.~G.}, \bibinfo{author}{Baldwin, K. G.~H.} \&
  \bibinfo{author}{Truscott, A.~G.}
\newblock
  \href{http://dx.doi.org/10.1126/science.1198481}{\bibinfo{title}{Direct
  {Measurement} of {Long}-{Range} {Third}-{Order} {Coherence} in
  {Bose}-{Einstein} {Condensates}}}.
\newblock \emph{\bibinfo{journal}{Science}} \textbf{\bibinfo{volume}{331}},
  \bibinfo{pages}{1046--1049} (\bibinfo{year}{2011}).

\bibitem{perrin_hanbury_2012}
\bibinfo{author}{Perrin, A.} \emph{et~al.}
\newblock \href{http://dx.doi.org/10.1038/nphys2212}{\bibinfo{title}{Hanbury
  {Brown} and {Twiss} correlations across the {Bose}–{Einstein} condensation
  threshold}}.
\newblock \emph{\bibinfo{journal}{Nature Physics}}
  \textbf{\bibinfo{volume}{8}}, \bibinfo{pages}{195--198}
  (\bibinfo{year}{2012}).

\bibitem{cayla_hanbury-brown_2020}
\bibinfo{author}{Cayla, H.} \emph{et~al.}
\newblock
  \href{http://dx.doi.org/10.1103/PhysRevLett.125.165301}{\bibinfo{title}{Hanbury-{Brown}
  and {Twiss} bunching of phonons and of the quantum depletion in a
  strongly-interacting {Bose} gas}}.
\newblock \emph{\bibinfo{journal}{Physical Review Letters}}
  \textbf{\bibinfo{volume}{125}}, \bibinfo{pages}{165301}
  (\bibinfo{year}{2020}).

\bibitem{raman_evidence_1999}
\bibinfo{author}{Raman, C.} \emph{et~al.}
\newblock
  \href{http://dx.doi.org/10.1103/PhysRevLett.83.2502}{\bibinfo{title}{Evidence
  for a {Critical} {Velocity} in a {Bose}-{Einstein} {Condensed} {Gas}}}.
\newblock \emph{\bibinfo{journal}{Physical Review Letters}}
  \textbf{\bibinfo{volume}{83}}, \bibinfo{pages}{2502--2505}
  (\bibinfo{year}{1999}).

\bibitem{weimer_critical_2015}
\bibinfo{author}{Weimer, W.} \emph{et~al.}
\newblock
  \href{http://dx.doi.org/10.1103/PhysRevLett.114.095301}{\bibinfo{title}{Critical
  {Velocity} in the {BEC}-{BCS} {Crossover}}}.
\newblock \emph{\bibinfo{journal}{Physical Review Letters}}
  \textbf{\bibinfo{volume}{114}}, \bibinfo{pages}{095301}
  (\bibinfo{year}{2015}).

\bibitem{kim_observation_2020}
\bibinfo{author}{Kim, J.~H.}, \bibinfo{author}{Hong, D.} \&
  \bibinfo{author}{Shin, Y.}
\newblock
  \href{http://dx.doi.org/10.1103/PhysRevA.101.061601}{\bibinfo{title}{Observation
  of two sound modes in a binary superfluid gas}}.
\newblock \emph{\bibinfo{journal}{Physical Review A}}
  \textbf{\bibinfo{volume}{101}}, \bibinfo{pages}{061601}
  (\bibinfo{year}{2020}).

\bibitem{recati_breaking_2019}
\bibinfo{author}{Recati, A.} \& \bibinfo{author}{Piazza, F.}
\newblock
  \href{http://dx.doi.org/10.1103/PhysRevB.99.064505}{\bibinfo{title}{Breaking
  of {Goldstone} modes in a two-component {Bose}-{Einstein} condensate}}.
\newblock \emph{\bibinfo{journal}{Physical Review B}}
  \textbf{\bibinfo{volume}{99}}, \bibinfo{pages}{064505}
  (\bibinfo{year}{2019}).

\bibitem{cominotti_observation_2021}
\bibinfo{author}{Cominotti, R.} \emph{et~al.}
\newblock \bibinfo{title}{Observation of {Massless} and {Massive} {Collective}
  {Excitations} with {Faraday} {Patterns} in a {Two}-{Component} {Superfluid}}.
\newblock \emph{\bibinfo{journal}{arXiv:2112.09880}}  (\bibinfo{year}{2021}).

\bibitem{betz_two-point_2011}
\bibinfo{author}{Betz, T.} \emph{et~al.}
\newblock
  \href{http://dx.doi.org/10.1103/PhysRevLett.106.020407}{\bibinfo{title}{Two-{Point}
  {Phase} {Correlations} of a {One}-{Dimensional} {Bosonic} {Josephson}
  {Junction}}}.
\newblock \emph{\bibinfo{journal}{Physical Review Letters}}
  \textbf{\bibinfo{volume}{106}}, \bibinfo{pages}{020407}
  (\bibinfo{year}{2011}).

\bibitem{bogolubov_theory_nodate}
\bibinfo{author}{Bogoliubov, N.}
\newblock \bibinfo{title}{On the theory of superfluidity}.
\newblock \emph{\bibinfo{journal}{J. Phys.}} \textbf{\bibinfo{volume}{11}},
  \bibinfo{pages}{23} (\bibinfo{year}{1947}).

\bibitem{kawaguchi_finite-temperature_2012}
\bibinfo{author}{Kawaguchi, Y.}, \bibinfo{author}{Phuc, N.~T.} \&
  \bibinfo{author}{Blakie, P.~B.}
\newblock
  \href{http://dx.doi.org/10.1103/PhysRevA.85.053611}{\bibinfo{title}{Finite-temperature
  phase diagram of a spin-1 {Bose} gas}}.
\newblock \emph{\bibinfo{journal}{Physical Review A}}
  \textbf{\bibinfo{volume}{85}}, \bibinfo{pages}{053611}
  (\bibinfo{year}{2012}).

\bibitem{blakie_solitons_2022}
\bibinfo{author}{Yu, X.} \& \bibinfo{author}{Blakie, P.~B.}
\newblock
  \href{http://dx.doi.org/10.1103/PhysRevLett.128.125301}{\bibinfo{title}{Propagating
  ferrodark solitons in a superfluid: Exact solutions and anomalous dynamics}}.
\newblock \emph{\bibinfo{journal}{Phys. Rev. Lett.}}
  \textbf{\bibinfo{volume}{128}}, \bibinfo{pages}{125301}
  (\bibinfo{year}{2022}).

\bibitem{prufer_observation_2018}
\bibinfo{author}{Prüfer, M.} \emph{et~al.}
\newblock
  \href{http://dx.doi.org/10.1038/s41586-018-0659-0}{\bibinfo{title}{Observation
  of universal dynamics in a spinor {Bose} gas far from equilibrium}}.
\newblock \emph{\bibinfo{journal}{Nature}} \textbf{\bibinfo{volume}{563}},
  \bibinfo{pages}{217--220} (\bibinfo{year}{2018}).

\bibitem{manz_two-point_2010}
\bibinfo{author}{Manz, S.} \emph{et~al.}
\newblock
  \href{http://dx.doi.org/10.1103/PhysRevA.81.031610}{\bibinfo{title}{Two-point
  density correlations of quasicondensates in free expansion}}.
\newblock \emph{\bibinfo{journal}{Physical Review A}}
  \textbf{\bibinfo{volume}{81}}, \bibinfo{pages}{031610}
  (\bibinfo{year}{2010}).

\bibitem{lannig_collisions_2020}
\bibinfo{author}{Lannig, S.} \emph{et~al.}
\newblock
  \href{http://dx.doi.org/10.1103/PhysRevLett.125.170401}{\bibinfo{title}{Collisions
  of three-component vector solitons in {Bose}-{Einstein} condensates}}.
\newblock \emph{\bibinfo{journal}{Physical Review Letters}}
  \textbf{\bibinfo{volume}{125}}, \bibinfo{pages}{170401}
  (\bibinfo{year}{2020}).

\bibitem{kunkel_spatially_2018}
\bibinfo{author}{Kunkel, P.} \emph{et~al.}
\newblock
  \href{http://dx.doi.org/10.1126/science.aao2254}{\bibinfo{title}{Spatially
  distributed multipartite entanglement enables {EPR} steering of atomic
  clouds}}.
\newblock \emph{\bibinfo{journal}{Science}} \textbf{\bibinfo{volume}{360}},
  \bibinfo{pages}{413--416} (\bibinfo{year}{2018}).

\end{thebibliography}
\setcounter{figure}{0}
\renewcommand{\figurename}{\textbf{Extended Data Figure}}

\newpage
\,

\begin{figure}
	\linespread{1}
	\centering
	\includegraphics[width = \columnwidth]{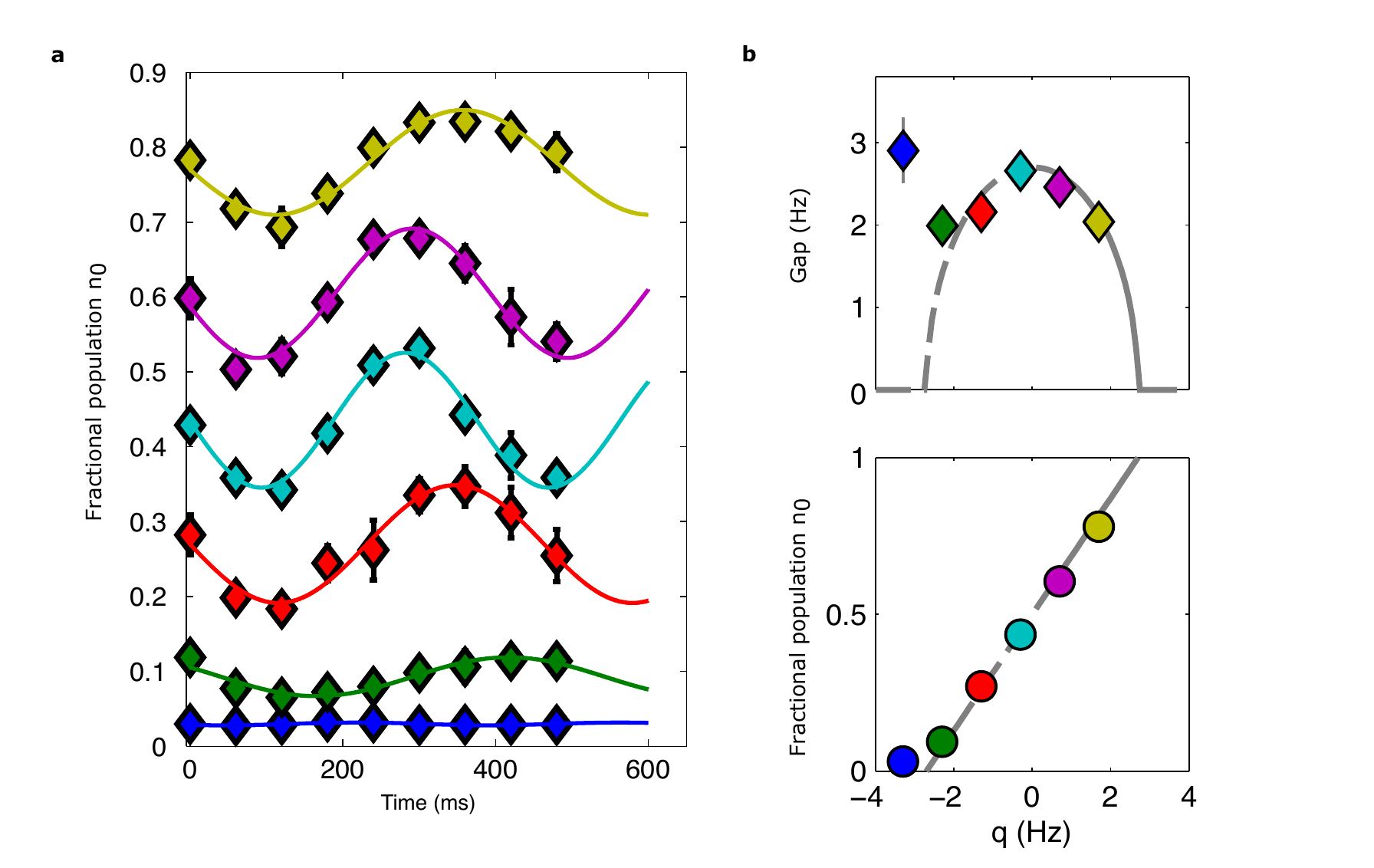} 
	\caption{\textbf{Measurement of the gap by observation of temporal oscillations of the $k=0$ mode. a}, We measure the gap of the quadratic spin mode by a global rotation of the spinor phase. We record the resulting oscillations of the fractional $m=0$ population as a function of evolution time after the rotation. We fit a sinusoidal function (solid line) to infer the frequency. \textbf{b}, Extracted oscillation frequency (diamonds) and mean value of the $m = 0$ population (circles). We compare to theoretical expectations for the easy-plane phase (solid lines; see Methods eq.\,(1)). The dashed line extrapolates the expectations to $q<0$ under the assumption of equal populations of $m = \pm 1$. For the theory curves we use $nc_1 = 1.3\,$Hz. }
	\label{Distribution}
\end{figure}

\begin{figure}
	\linespread{1}
	\centering
	\includegraphics[width = \columnwidth]{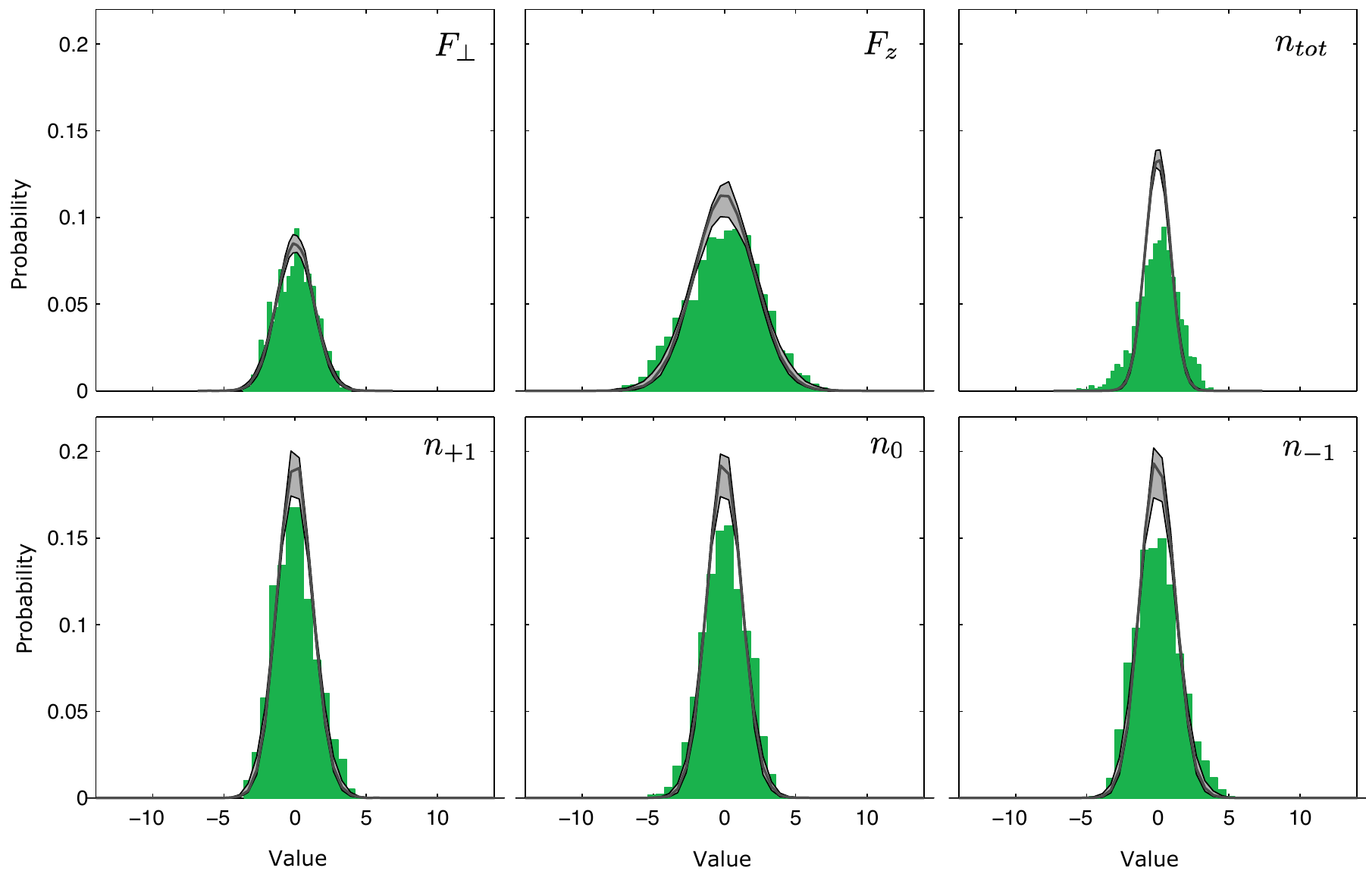} 
	\caption{\textbf{Histograms of local observables in the thermalized state.} Histograms obtained from evaluating the local observations of the experimental data presented in Fig. 4 (green bars). Here, each local observable is normalized to the square-root of the local mean of the total atom number. On top we display theoretical estimates from 1000 samples generated according to thermal Bogoliubov theory with parameters as in Fig. 4 (grey line; grey band indicates 68\% confidence interval including statistical and systematic uncertainties). The mean value of each histogram is subtracted. For details on the sampling procedure see Methods.}
	\label{Distribution}
\end{figure}

\begin{figure}
	\linespread{1}
	\centering
	\includegraphics[width = \columnwidth]{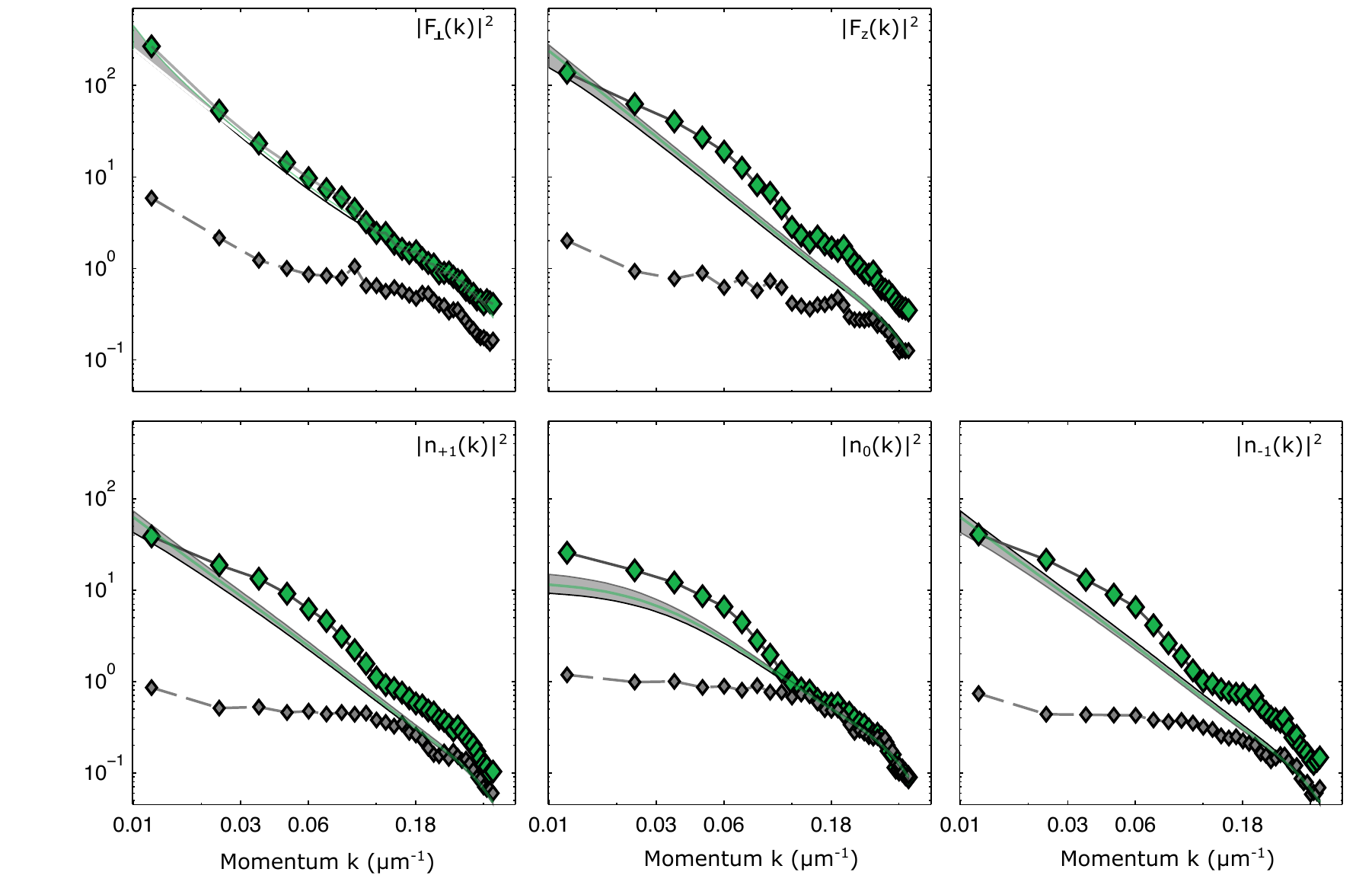} 
	\caption{\textbf{Structure factor  close to $q=0$.} We show experimental power spectra of different spin and density degrees of freedom close to $q=0$ (green diamonds). The grey diamonds represent the fluctuations of a coherent spin state with comparable atom numbers. We compare to thermal Bogoliubov theory predictions for the same parameters as displayed in Fig. 4 but with $q=0$ (green line; grey band indicates 68\% confidence interval of statistical and systematic uncertainties). {Experimentally, we find that for momenta in the range of $0.02\mu\text{m}^{-1}$ to $0.1\mu\text{m}^{-1}$ the fluctuations are higher than for the thermal predictions for all observables (except the transversal spin $F_\perp$). The length scale of these fluctuations is in accordance with observable localized long-lived non-linear excitations which are not present in the thermalized data of Fig.\,4. }}
	\label{Distribution}
\end{figure}

\end{document}